\documentclass[sigconf]{acmart}
\AtBeginDocument{%
  \providecommand\BibTeX{{%
    \normalfont B\kern-0.5em{\scshape i\kern-0.25em b}\kern-0.8em\TeX}}}

\setcopyright{acmcopyright}
\copyrightyear{2026}
\acmYear{2026}

\acmConference[DIS '26]{Designing Interactive Systems Conference}{June 13--17, 2026}{Singapore, Singapore}
\acmBooktitle{Designing Interactive Systems Conference (DIS '26), June 13--17, 2026, Singapore, Singapore}

\usepackage{_macros}
\usepackage{xcolor}
\usepackage{listings}
\usepackage[ruled,linesnumbered]{algorithm2e}
\usepackage[normalem]{ulem}

\newcommand{\revision}[1]{\leavevmode{\color{blue}{#1}}}
\newcommand{\remark}[1]{\leavevmode{\color{red}{! #1}}}
\newcommand{\removed}[1]{\leavevmode{\color{red}{\st{#1}}}}

\def \cleanversion{} 
\ifx\cleanversion\undefined
\else
 \renewcommand{\remark}[1]{\iffalse #1 \fi} 
 \renewcommand{\removed}[1]{\iffalse #1 \fi} 
 \renewcommand{\revision}[1]{#1}
\fi

\newcommand{\name}{MindTrellis}

\renewcommand{\wsicon}[1]{\raisebox{-0.3\height}{\includegraphics[height=1.2em]{image/icons_screenshots/#1.png}}}

\makeatletter
\renewcommand{\@seccntformat}[1]{\textbf{\csname the#1\endcsname}\quad}
\makeatother

\begin{document}

\title[\name]{MindTrellis: Co-Creating Knowledge Structures with AI through Interactive Visual Exploration}

\author{Xiang Li}
\authornote{Both authors contributed equally to this research.}
\affiliation{%
  \institution{University of Waterloo}
  \city{Waterloo}
  \state{Ontario}
  \country{Canada}}
\email{x247li@uwaterloo.ca}

\author{Cara Li}
\authornotemark[1]
\affiliation{%
  \institution{University of Waterloo}
  \city{Waterloo}
  \state{Ontario}
  \country{Canada}}
\email{cy5li@uwaterloo.ca}

\author{Emily Kuang}
\affiliation{%
  \institution{York University}
  \city{Toronto}
  \state{Ontario}
  \country{Canada}}
\email{ekuang@yorku.ca}

\author{Can Liu}
\affiliation{%
  \institution{Nanyang Technological University}
  \city{Singapore}
  \country{Singapore}}
\email{can.liu@ntu.edu.sg}

\author{Jian Zhao}
\affiliation{%
  \institution{University of Waterloo}
  \city{Waterloo}
  \state{Ontario}
  \country{Canada}}
\email{jianzhao@uwaterloo.ca}

\renewcommand{\shortauthors}{Li et al.}

\begin{abstract}
Synthesizing information from multiple documents into structured understanding is inherently iterative, yet current approaches provide limited support.
LLM-based systems let users query information but produce structures that users cannot reshape; manual tools like mind maps offer full control but lack intelligent assistance; \revision{and commercial tools have begun combining retrieval with user contribution, but not within a unified visual knowledge structure.}
We present \name{}, an interactive visual system that addresses this gap by letting users and AI collaboratively build a knowledge graph \revision{combining document-derived and user-contributed knowledge}.
Users can query the graph to retrieve document-grounded information, and contribute new concepts, relationships, and hierarchical organization to reflect their developing understanding.
\revision{A multi-agent pipeline coordinates intent disambiguation, knowledge placement, and coherence maintenance across both pathways.}
\revision{
In a controlled study where 12 participants created slide decks, \name{} outperformed a retrieval-only baseline in knowledge organization and cognitive load, with participants valuing progressive graph expansion and the ability to integrate their own insights.

}

\remark{change log: (1) Acknowledged commercial tools in landscape framing, consistent with revised Introduction. (2) Replaced ``dynamic knowledge graph'' with co-created KG terminology. (3) Added multi-agent pipeline sentence (contribution 2). (4) Revise the claim to match actual study measures.}

\end{abstract}




\begin{teaserfigure}
  \includegraphics[width=\textwidth]{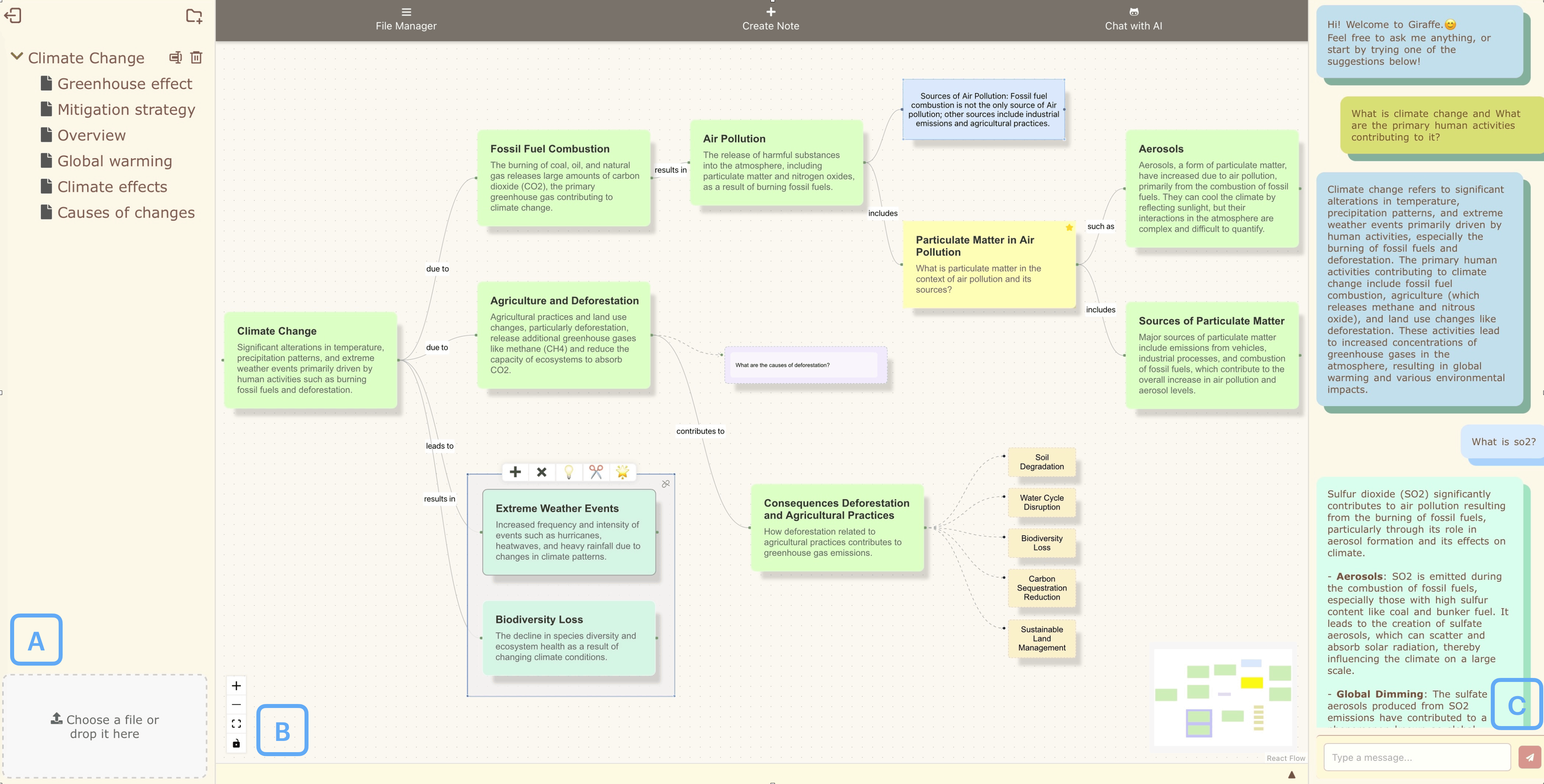} 
  \vspace{-6mm}
  \caption{
\name{} is an interactive visual system that enables human-AI collaborative knowledge construction, where users can both query information and contribute insights to an evolving knowledge graph.
(a) File Manager -- allows users to upload and organize documents.
(b) Knowledge Canvas -- displays the knowledge graph where users can explore and directly manipulate nodes and relationships.
(c) Chat Panel -- enables users to query information, request expansions, and issue modification commands through natural language. }
  \label{fig:teaser}
  \Description{ Screenshot of the \name{} interface with three panels: a file manager listing uploaded documents on the left, a central canvas showing a hierarchical node-link knowledge graph about climate change with labeled edges connecting concepts, and a chat panel on the right displaying user queries and AI responses.}
\end{teaserfigure}

\maketitle

\section{Introduction}

As digital resources continue to expand, knowledge workers face increasing challenges in understanding and synthesizing information from multiple sources.
Developing a coherent understanding is inherently an iterative process: users explore information, form mental models, refine their understanding, and reorganize their knowledge as they learn~\cite{pirolli2005sensemaking}.
However, current tools provide limited support for this evolving process, creating a gap between how human understanding develops and how systems facilitate knowledge construction.
Traditional visualization tools based on mind maps~\cite{buzan2006use} and concept maps~\cite{novak1984learning} allow users to construct and modify knowledge representations, offering full control over structure.
However, they lack sufficient intelligent assistance, largely relying on users' manual operations, which makes the process tedious and cognitively demanding when dealing with complex, multi-source information.
In recent years, researchers have developed automatic knowledge graph construction methods that use extraction algorithms to identify entities and relationships from text corpora~\cite{hogan2021knowledge}, but this process involves minimal human participation and produces structures optimized for machine consumption rather than human understanding.
Since the emergence of Large Language Models (LLMs), retrieval-augmented generation (RAG) systems~\cite{lewis2021rag} have enabled users to query knowledge bases through natural language.
\revision{Recent systems such as Graphologue~\cite{graphologue}, Sensecape~\cite{suh_sensecape_2023}, Selenite~\cite{liu2024selenite}, and Luminate~\cite{suh2024luminate} further integrate LLMs with graphical representations to support visual exploration of complex information.}
Yet these systems treat the underlying knowledge structure as \textit{largely static}: users can retrieve information and receive AI-generated visualizations, 
but cannot add their own concepts, modify relationships, or reorganize how the structure is arranged.
%
\revision{The broader pattern of users shaping the same representations they consume from is well established in interactive systems~\cite{hempel2019sketch, ye2020penrose}, and commercial tools such as NotebookLM and Notion AI already let users both query and contribute to underlying knowledge stores.
These commercial tools, however, distribute retrieval and contribution across separate panes or database views rather than integrating them within a single visual knowledge structure.}

\revision{Multi-document knowledge construction is fundamentally iterative~\cite{russell1993cost}: users externalize their developing mental models~\cite{kirsh2017thinking}, test relationships between concepts~\cite{novak1984learning, canas2005concept}, and continuously reorganize their representations as understanding deepens~\cite{pirolli2005sensemaking, binks2022representational}.
Research on knowledge visualization suggests that spatial representations reduce cognitive effort~\cite{larkin1987diagram} and that the cognitive value of working with multiple representations lies in the transformations between them~\cite{binks2022representational}.
A system designed to support this process should therefore integrate AI-assisted retrieval and user contribution within a shared visual structure, so that users can see how new content relates to existing knowledge and reorganize the whole as their understanding evolves.}

\revision{We present \textit{\name{}}, an interactive visual system for multi-document knowledge construction whose design centers on a visual knowledge graph as the primary shared artifact (Fig.~\ref{fig:teaser}).}
In one direction, users can \textit{query} the knowledge base to retrieve information grounded in their uploaded documents.
In the other direction, users can \textit{contribute} to the knowledge structure by adding new concepts, modifying relationships, and reorganizing the hierarchy to reflect their developing understanding; the system simultaneously synchronizes these contributions to the underlying knowledge base to maintain consistency between the visual representation and stored knowledge.
\revision{The resulting representation is what we call a \textit{co-created knowledge graph}: a hybrid of document-derived content and user-contributed insights that evolves through the collaboration between human understanding and AI assistance.}
 
\remark{change log (paragraphs 1--3): (1) Removed ``fundamentally unidirectional'' and ``What remains missing is a contribution pathway'' per shepherd's instruction to de-emphasize novelty claims. (2) Added landscape acknowledgment (Sketch-n-Sketch, Penrose, NotebookLM, Notion AI) per shepherd's instruction to acknowledge existing bidirectional systems; commercial tools named without footnotes (footnoted in Section~2.2). (3) Expanded LLM systems list to include Selenite and Luminate, reflecting revised Related Work. (4) Replaced gap-framing paragraph with theoretical motivation per Russell et al.; distributed citations across cognitive activities. (5) Reframed MindTrellis introduction to lead with design contribution rather than gap-filling; rewrote co-created KG definition to emphasize hybrid nature, consistent with Section~2.3.}

However, enabling this co-creation introduces new technical challenges.
User input is inherently ambiguous: a query like \qt{Expand the effects of deforestation on causing global warming} could be a request for information retrieval, a command to expand the knowledge structure, or both simultaneously.
When users contribute new knowledge, the system needs to determine where this information belongs within the existing taxonomy, which requires understanding both the new content and the current graph topology.
As the knowledge graph evolves through user contributions, the system must also maintain coherence and consistency, avoiding contradictions and redundancies.
To address these challenges, we developed a \textit{multi-agent pipeline} in \name{} where specialized components collaborate to support seamless bidirectional interaction: classifying user intent, routing inputs to appropriate processing pipelines, handling retrieval at multiple granularity levels, and executing structural modifications while maintaining graph coherence.

Based on the co-created knowledge graph and the multi-agent system, \name{} supports two complementary interactions for controlling and modifying the knowledge graph, within a unified environment.
Through the chat interface, users can issue natural language commands to query information, add nodes, or restructure relationships. 
Through the visual canvas, users can directly manipulate~\cite{hutchins1985direct} the graph by clicking to select nodes, dragging to reposition elements, and using toolbar controls to expand topics or create connections.

To evaluate \name{}, we conducted a controlled user study with 12 participants comparing our system against a baseline system that combines a RAG interface with automatically generated graph visualizations. 
\revision{The baseline supports the query direction, where users can ask questions and view document-grounded answers as node diagrams, but does not allow users to modify the generated structure or add their own concepts.}
Participants rated \name{} significantly higher on knowledge organization effectiveness and reported lower frustration when exploring unfamiliar topics. 
Qualitative feedback highlighted that progressive expansion of the knowledge graph, where new nodes branch from existing ones rather than appearing all at once, reduced participants' cognitive overload and helped maintain a coherent mental model. 
Participants also valued the ability to integrate their own insights into the evolving structure.
These findings suggest that enabling users to actively shape knowledge representations, rather than passively consuming system-generated structures, can meaningfully improve outcomes in information exploration tasks.

In summary, our contributions in this paper include:
\begin{itemize}
\item \revision{The design and implementation of \name{}, an interactive system for knowledge construction from multi-document sources, featuring a visual knowledge graph where document-derived and user-contributed knowledge coexist, manipulable through both natural language and direct manipulation.}
\item A multi-agent pipeline that addresses the technical challenges of intent parsing, knowledge placement, and coherence maintenance, validated through quantitative evaluation.
\item Findings from a controlled user study demonstrating that \name{} improves knowledge organization effectiveness and reduces cognitive load compared to retrieval-only baselines.
\end{itemize}

\remark{change log (paragraphs 4--7): (1) Replaced overclaim about single-prompt models with forward reference to Section~\ref{sec:system} and Section~\ref{sec:pipeline_eval} per revision plan. (2) Rewrote contribution 1 to foreground system design rather than ``bidirectional interaction'' as novelty, per the shepherd's instruction to emphasize design contribution. (3) Fixed grammar in contribution 2 and the study results paragraph.}
\section{Related Work} \label{sec:related_work}

\subsection{\revision{Knowledge Visualization and Externalization}}
 
As both digital and physical resources continue to expand, people increasingly recognize the value of visualizing knowledge~\cite{eppler2008knowledge,meyer2010} to aid in learning and understanding complex information~\cite{safar2014mind,canas2005concept}. 
Knowledge visualization provides a structured way to externalize thoughts, which is particularly useful for understanding complex or multi-faceted subjects~\cite{wang2011guest}. 
Different forms of visualization, especially graph-based representations, can significantly enhance critical thinking~\cite{kasumu2022concept,machado2020concept} and comprehension~\cite{dias2011concept,rassaei2019effects}, which includes concept maps~\cite{davenport1998working,o2002knowledge,lee2012knowledge} and mind maps~\cite{erdem2017mind,wickramasinghe2011effectiveness}. 
Overall, these visualization techniques support memory retention and comprehension as well as facilitate deeper cognitive engagement and a more structured understanding of complex topics.
\revision{However, not all visual forms are equally effective, and the choice of representation can shape what people notice, how they reason, and what knowledge they construct.}
 
\revision{Carneiro et al.~\cite{carneiro2021text} compared textual and graphical representations for argument analysis and found that graph-based interfaces better support reasoning over non-linear argument structures, where relationships between claims are difficult to convey through sequential text.
Binks et al.~\cite{binks2022representational} studied how users move between map-based and textual representations during essay writing, finding that the cognitive value lies not within any single representation but in the \textit{representational transformations} between them---the process of re-expressing and reorganizing knowledge across formats.
Their characterization of transformation properties (cardinality, explicitness, and representation type change) suggests that systems for knowledge work should facilitate fluid movement between complementary representations.
Larkin and Simon~\cite{larkin1987diagram} offered an earlier theoretical account of these effects, showing that spatial and diagrammatic representations can reduce cognitive effort by supporting perceptual inference and reducing search compared to informationally equivalent text.}
\revision{The above work motivates us to design \name{} that goes beyond static visualization to support interactive construction and manipulation of knowledge structures.}
 
\remark{change log: (1) Renamed subsection to ``Knowledge Visualization and Externalization.'' (2) Added Carneiro et al. (VL/HCC 2021) and Binks et al. (IJHCS 2022) per R4 recommendation to incorporate these references; added Larkin \& Simon (1987) for foundational grounding. (3) Removed Graphologue/Sensecape paragraph (moved to 2.2) and ``most systems...MindTrellis addresses'' gap claim per the shepherd's instruction to de-emphasize novelty. (4) Added a transitional sentence at the end of the first paragraph to improve flow into new content.}

\revision{
\subsection{Bidirectional Interaction Between Representations}

Bidirectional interaction between coupled representations is a well-established design pattern in interactive systems and programming languages~\cite{foster2007lenses}.
In such systems, users can shape the same artifact they consume from, and the system keeps both views synchronized.
Sketch-n-Sketch~\cite{hempel2019sketch} enables users to write programs that generate SVG graphics and then directly manipulate the rendered output; the system infers corresponding program updates through trace-based synthesis, keeping code and visual artifact synchronized.
Penrose~\cite{ye2020penrose} maps mathematical notation to diagrams through a trio of domain-specific languages and constraint-based optimization, translating formal specifications into visual layouts.
B2~\cite{wu2020b2} bridges code and interactive visualizations in computational notebooks by treating data queries as a shared intermediate representation, so that interactions with a chart reify as code and vice versa.
The pattern extends to other domains as well: Cascaval et al.~\cite{cascaval2022differentiable} apply bidirectional editing to parametric CAD programs, where users can manipulate geometry directly and the system solves an inverse problem to update program parameters.
In each of these systems, two views of a single underlying artifact are kept in sync through deterministic or constraint-based mechanisms---a pattern we refer to as \textit{representational synchronization}.
The mapping is mechanical rather than interpretive, the system does not need to infer what the user intends, because the editing modality is spatially explicit (a code panel versus an output canvas) and the artifact on both sides is the same.

As bidirectional interaction extends from structured authoring and design tools into constructing knowledge structures from multiple documents, the nature of the pattern changes.
The coupled representations are no longer two views of one artifact but two distinct kinds of content: source documents and an evolving knowledge graph.
Users contribute concepts that may not appear in any source document~\cite{pirolli2005sensemaking}, so the graph becomes a hybrid of document-derived and user-contributed knowledge---a structure richer than either source alone.
Because retrieval and contribution share the same natural language input channel, the system must determine which activity a given input represents.
A statement like \qt{Expand the effects of deforestation on the cause global warming} could be a question seeking retrieval or a suggestion to add a new concept to the graph.
Shahriari et al.~\cite{shahriari2025nl} studied natural language interaction for editing visual knowledge graphs and found that distinguishing between edit intent and information queries is a central design challenge.
The transformation between representations is therefore AI-mediated and semantically interpretive, requiring the system to classify user intent before determining how to act.
Intent disambiguation, coherent placement within an evolving structure, and maintenance of a hybrid artifact are requirements specific to knowledge-level bidirectionality; they do not arise in the representational synchronization paradigm, where editing modalities are structurally partitioned and mappings are deterministic.
}

\remark{change log: (1) New subsection replacing original 2.2 (``LLM Agents in Information Seeking''), which was condensed into a paragraph in 2.3. (2) Added Sketch-n-Sketch (UIST 2019), Penrose (SIGGRAPH 2020), B2 (UIST 2020), and Cascaval et al. (Eurographics 2022) per shepherd's instruction to include bidirectional editing systems and contextualize MindTrellis relative to them. Foster et al. (TOPLAS 2007) cited for foundational lens theory. (3) Added Shahriari et al. (K-CAP 2025) on NL editing of visual knowledge graphs to support the intent ambiguity challenge. (4) Framed as a progression from representational synchronization to knowledge-level bidirectionality per shepherd's instruction to qualify novelty claims.}
\subsection{Interactive Systems for Knowledge Management}

Traditional knowledge management tools primarily support static representations, requiring users to manually create, organize, and update all connections and relationships~\cite{davenport1998working}.  
\revision{However, manual construction becomes cognitively demanding as datasets grow larger and more complex, and the resulting structures often lack the visual scaffolding needed to help users organize and make sense of information across sources~\cite{eppler2008knowledge}.}
A parallel line of research has focused on automatic knowledge graph construction, where systems extract entities and relationships from text corpora using named entity recognition, relation extraction, and knowledge base population techniques~\cite{ji2022survey}.
While these approaches can process large volumes of documents efficiently, they are designed primarily for machine consumption---producing structured databases optimized for computational queries---rather than human \revision{knowledge building}.
The resulting knowledge graphs reflect algorithmic decisions with no mechanism for incorporating human insight or adapting to individual users' evolving understanding.
Furthermore, errors in automatic extraction propagate through the structure without opportunities for user correction or refinement.

\revision{The progression from manual authoring to automatic construction to LLM-augmented exploration reflects a broader trajectory in knowledge representation systems~\cite{ji2022survey, pirolli2005sensemaking}.
Manual visualization tools offer full user control but no intelligent assistance~\cite{davenport1998working, novak1984learning}.
Automatic knowledge graph systems leverage computational power but exclude human participation, producing structures that may not align with how users need to understand information~\cite{mitchell2018nell, dong2024knowledge_vault}.
Recent LLM-augmented systems generate structured visualizations from LLM outputs but provide limited support for users to reshape the underlying knowledge structure.}

Graphologue~\cite{graphologue} converts LLM text responses into interactive node-link diagrams, enhancing comprehension of complex answers, but the generated structure remains a read-only output that users cannot refine to reflect their evolving understanding.
Sensecape~\cite{suh_sensecape_2023} supports multilevel exploration through hierarchical diagrams at different abstraction levels, yet the knowledge structure itself is system-determined.
\revision{Selenite~\cite{liu2024selenite} addresses the cold-start problem in sensemaking by generating comprehensive overviews of options and criteria from LLMs, scaffolding exploration of unfamiliar domains.
Luminate~\cite{suh2024luminate} takes a different approach by structuring the \textit{design space} of LLM outputs, enabling users to explore diverse responses along meaningful dimensions rather than converging on a single answer.
D\"{u}ck et al.~\cite{duck2025needles} support claim retrieval in large document corpora through multiple exploration pathways, combining keyword search with hypothesis-driven retrieval and consistency checking.
Each of these systems advances a specific aspect of LLM-augmented knowledge exploration, yet the resulting representations are generated by the system and not persistently editable by users---users can navigate and query but cannot contribute their own knowledge to an evolving shared structure.}
 
\revision{Commercial tools have also begun incorporating AI capabilities for knowledge work.
NotebookLM\footnote{\url{https://notebooklm.google.com}} supports document-grounded question answering over user-uploaded sources and recently added AI-generated mind maps that visualize connections between source concepts.
Notion AI\footnote{\url{https://www.notion.com/product/ai}} enables retrieval and content generation across a workspace of pages and databases.
Miro AI\footnote{\url{https://miro.com}} offers AI-assisted diagram generation on a collaborative visual canvas, while Excalidraw\footnote{\url{https://excalidraw.com}} and Lucidchart\footnote{\url{https://www.lucidchart.com}} support text-to-diagram conversion with AI features.
Several of these tools already support forms of knowledge-level bidirectionality.
NotebookLM and Notion AI, for example, let users both query and contribute to underlying knowledge stores.
However, NotebookLM's core interaction is document-grounded question answering; mind maps are a secondary output of that process rather than the central artifact users build upon.
Notion AI, similarly, operates entirely within a page-and-database structure with no visual knowledge graph at all.
Contributions and retrievals thus remain distributed across separate views rather than integrated within a single visual structure.
Research on knowledge visualization suggests this separation has cognitive consequences: Carneiro et al.~\cite{carneiro2021text} found that graph-based interfaces better support reasoning over non-linear structures than equivalent textual representations, Larkin and Simon~\cite{larkin1987diagram} showed that spatial representations reduce cognitive effort by supporting perceptual inference, and Binks et al.~\cite{binks2022representational} found that the cognitive value of knowledge representations lies in the transformations between them.
Together, these findings suggest that integrating retrieval and contribution within a single visual knowledge structure may support reasoning that separate-pane designs do not afford.
In such a structure, users can see how new content relates to existing knowledge and reorganize the whole.

}

On the technical side, Retrieval-Augmented Generation (RAG)~\cite{lewis2021rag} has become a foundational approach for grounding LLM responses in external knowledge bases, combining generative capabilities with document retrieval to improve factual accuracy~\cite{shuster2021retri}.
Subsequent work has extended RAG through hierarchical retrieval strategies such as RAPTOR~\cite{sarthi2024raptor}, which recursively clusters and summarizes text chunks to enable retrieval at varying levels of granularity.
Multi-agent architectures~\cite{wu2023autog} have further expanded the capabilities of LLM systems by coordinating specialized components for complex tasks.
\revision{However, existing agent architectures are designed primarily for task completion rather than knowledge structure evolution~\cite{sumers2024cognitive, wang2024survey}---they treat the knowledge base as a static resource to be queried, not a shared artifact to be co-constructed.
\name{} addresses the challenges with a multi-agent pipeline, enabling knowledge construction to be treated as collaborative rather than system-determined.}
 
\revision{Across the research and commercial landscape surveyed above, research systems advance retrieval, visualization, or AI-assisted exploration but produce structures users cannot reshape; commercial tools have begun supporting knowledge-level bidirectionality but distribute retrieval and contribution across separate views.
The theoretical and empirical evidence on knowledge visualization suggests that spatially integrating both activities within a unified visual structure may 
yield cognitive benefits that separate-pane designs do not afford.
\name{} pursues this direction, enabling users and AI to co-construct an evolving knowledge graph where both querying and contributing operate on the same structure grounded in source documents.}
 
\remark{change log: (1) Softened ``three eras'' framing per shepherd's instruction to qualify novelty claims --- replaced ``strict separation'' with ``limited support.'' (2) Added Selenite (CHI 2024) and Luminate (CHI 2024) per R3 recommendation; added D\"{u}ck et al. (CHI 2025) per committee recommendation. Graphologue and Sensecape moved here from 2.1 and reframed as part of a broader landscape rather than as the sole motivation. (3) Added commercial tools paragraph (NotebookLM, Notion AI, Miro AI, Excalidraw, Lucidchart) per shepherd's instruction to contextualize with commercial tools and R2's concern about existing systems. Honestly acknowledged capability overlap. (4) Condensed RAG/multi-agent content from original Section 2.2 into one paragraph; removed ``cascading forgetfulness'' per R2 feedback on dated claims. (5) Replaced ``What remains missing'' gap claim with observational closing per shepherd's instruction to de-emphasize novelty.}

\section{Designing \name{}} 
To understand user needs for knowledge building from multiple sources, we conducted a formative study that informed six key challenges; we then derived design goals that guided the development of \name{}.

\subsection{Formative Study}

We recruited six graduate students with diverse academic backgrounds (three female and three male) in computer science (two), neuroscience (two), and human-computer interaction (two).
The study included a 20-minute knowledge exploration task where participants studied three psychology documents they had no prior familiarity with, followed by a 10-question open-book quiz testing their comprehension of the material.
Participants first explored using only the three source documents without any AI assistance, then explored the same materials using a linear chatbot interface powered by a RAPTOR retriever (the same retriever used in our main user study).
After completing both phases, we conducted semi-structured interviews covering their information-seeking strategies, challenges encountered, and preferences for tool support.
The interviews were audio-recorded and transcribed for analysis.
Two authors independently conducted initial open coding of the transcripts, then collaboratively aligned codes and constructed themes.
We identified six key challenges that motivated our design.

\textbf{C1: Navigating and synthesizing multiple documents is cognitively demanding.}
When exploring without AI assistance, participants found it tedious to locate and integrate information scattered across three separate documents.
P6 described the experience as \textit{``pretty tedious... I was slightly annoyed because I had to keep switching back and forth.''}
P2 noted the difficulty of tracking information across sources: \textit{``I found it in one note but couldn't find the corresponding keyword''} in another.
The cognitive burden of manually navigating, cross-referencing, and synthesizing content left participants fatigued and prone to missing relevant connections, motivating the need for AI-assisted tools.

\textbf{C2: Connecting related concepts across sources remains difficult.}
Even with AI assistance, participants struggled to recognize when the same concept appeared across documents with different terminology.
P3 observed that \textit{``the wording is different, maybe they use different terminology to describe the same thing.''}
While the RAG-based chatbot could retrieve relevant passages, it did not explicitly surface connections across sources or help users reconcile different framings of the same concept.
The burden of integration remained on users in both phases, who had to manually identify that two differently-worded passages referred to the same underlying idea.

\textbf{C3: Retrieved content lacks structural organization for knowledge building.}
While the RAG-based chatbot reduced navigation burden by retrieving relevant information, participants found that linear text responses made it difficult to understand relationships between concepts.
P1 requested \textit{``a mind map''} because \textit{``a pile of text is not as intuitive as a diagram.''}
P5 similarly suggested \textit{``a mind map or matrix map to visualize the information,''} and P3 wanted \textit{``an outline with a list of keywords that lets me navigate back to the original content.''}
These comments reveal a mismatch between how conversational LLM interfaces present information and how users naturally organize knowledge: users need to see not only individual facts but also how concepts relate hierarchically and semantically.

\textbf{C4: LLM-generated content poses trust and verification concerns.}
While participants appreciated RAG-assisted retrieval efficiency, many hesitated to trust responses without verification.
P4 articulated this tension: \textit{``the problem with AI tools is for academia. I cannot see from where that content is coming.''}
P3 noted that for important tasks, they \textit{``might even want to use backtracking for every question to go back to the source and make sure it didn't hallucinate.''}
P6 felt \textit{``less confident compared to the notes. I'm worried if there will be hallucination.''}
Without clear provenance linking LLM-generated content to source materials, users face an uncomfortable choice between efficiency (accepting AI output) and accuracy (manually verifying everything).

\textbf{C5: Exploration does not culminate in an evolving knowledge structure.}
When using the chatbot, participants found that interactions were transient; each query produced an independent response with no persistent structure building over time.
P6 noted that the chatbot \textit{``doesn't remember the conversation history''} and wished follow-up questions could build upon previous context.
P3 described a similar issue with retention: after receiving responses from the chatbot, P3 reflected \textit{``I don't remember anything,''} suggesting isolated responses fail to support lasting knowledge construction.
Effective knowledge building is cumulative: users progressively expand understanding by building on prior knowledge and constructing comprehensive mental models.
When each interaction exists in isolation, users cannot see their exploration history or develop a coherent overall structure.

\textbf{C6: AI-generated structures are static and non-modifiable.}
Even when LLM provides organized output, participants wanted to reshape it according to their own understanding, but found existing systems did not support such customization.
P5 emphasized that \textit{``this kind of exploration is very personal''} and \textit{``the logic follows how you understand the thing,''} expressing desire to \textit{``customize''} how information is structured.
P1 similarly wished to \textit{``export a document... then iterate and update''} the content.
Knowledge building is inherently personal and iterative: users develop understanding by actively reorganizing and refining information structures.
Systems producing static, non-editable output prevent users from engaging in this essential process.

\subsection{Design Goals} \label{sec:design_goals}

Drawing from the challenges identified in our formative study (C1--C6), we established the following design goals to guide the development of \name{}.

\textbf{D1: Represent multi-sourced knowledge through a co-created, visually-structured graph (C1, C2, C3).}
A co-created knowledge graph combines the hierarchical organization of mind maps with the semantic relationships of concept maps~\cite{sarrafzadeh2016knowledge}, enabling users to see how concepts connect in an intuitive, layered format.
This representation makes structural relationships explicit and visible, reducing the cognitive effort of navigating multiple sources (C1) and understanding conceptual connections (C3).
Integrating information from multiple sources into a unified visual structure also helps users identify cross-document connections that might otherwise remain hidden (C2).
The knowledge graph is co-created in the sense that AI provides intelligent assistance in organizing and connecting information, while users retain the ability to shape and refine the structure according to their understanding.

\textbf{D2: Enable bidirectional interaction for cumulative knowledge building (C5, C6).}
Bidirectional interaction allows users to both query from and contribute to the knowledge structure~\cite{al2020using, al2023emerging}.
In the query direction, users retrieve information grounded in their documents, with responses organized within the evolving knowledge graph.
In the contribution direction, users actively shape the structure by adding concepts, modifying relationships, and reorganizing hierarchies to reflect their developing understanding.
This transforms users from passive consumers into active co-creators.
Contributions persist and accumulate~\cite{nevcasky2022interactive}, allowing the knowledge graph to evolve continuously as users explore and refine their mental models (C5).
Supporting modification of AI-generated structures enables the personal, iterative refinement essential to deep knowledge building (C6).

\textbf{D3: Support flexible interaction with transparent provenance (C4).}
Two complementary interaction modes accommodate diverse preferences and task demands~\cite{cohen1989synergistic}.
Natural language interaction through a chat interface enables conversational queries and structural modification commands.
Direct manipulation on a visual canvas~\cite{hutchins1985direct, miller1999improving} provides precise control through clicking, dragging, and toolbar operations.
Natural language is efficient for complex queries and bulk operations; direct manipulation offers spatial understanding and fine-grained control.
Regardless of interaction mode, transparent provenance allows users to trace any information back to its source document, directly addressing trust concerns by supporting verification when needed (C4).

\section{\name{}} \label{sec:system}

\begin{figure*}[t]
    \centering
    \includegraphics[width=\linewidth, page=1]{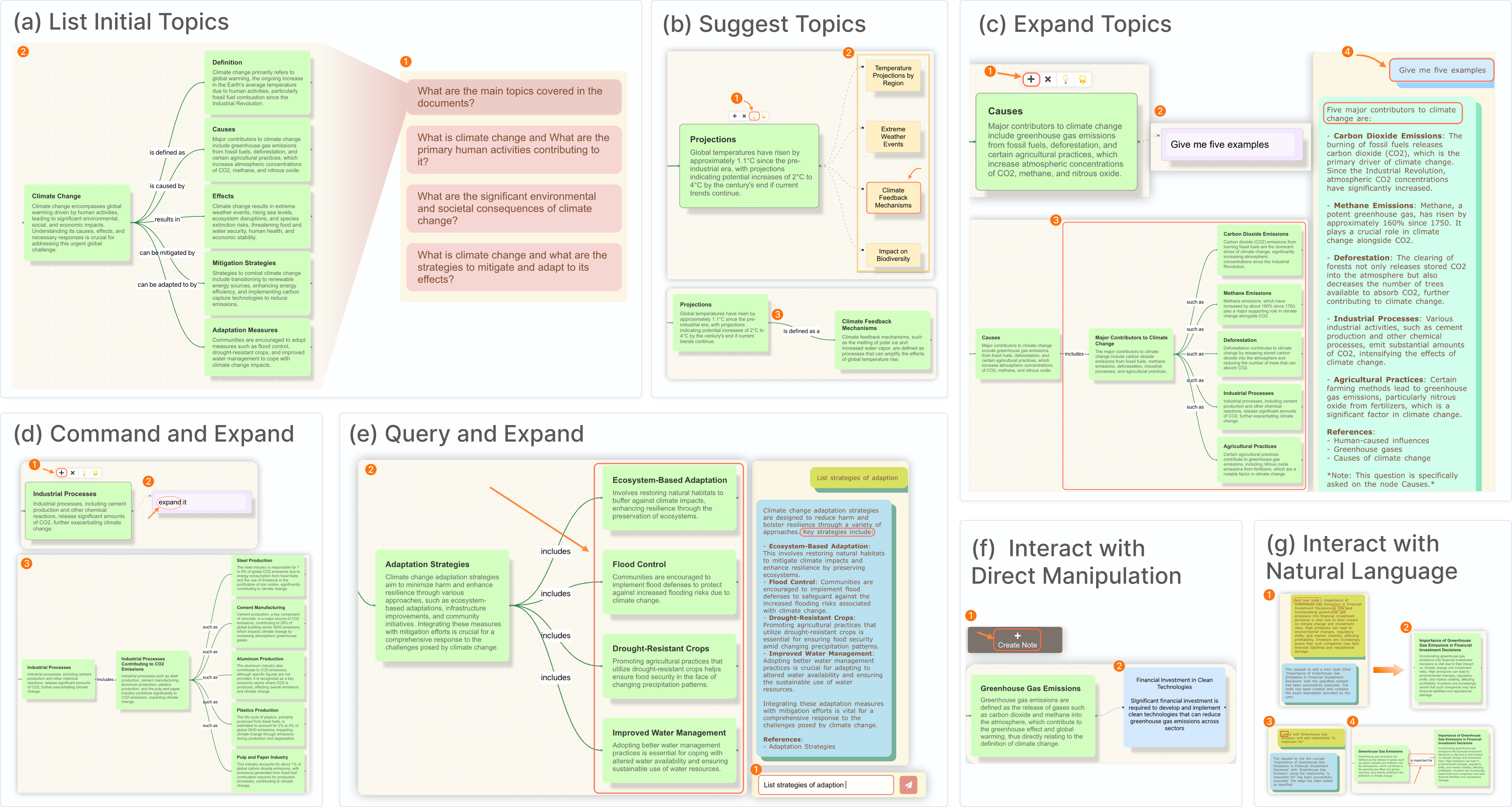}
    \vspace{-6mm}
    \caption{ 
    Key functionalities of \name{} demonstrated through the user scenario. (a) Initial exploration generating a hierarchical overview of climate change concepts. (b) System-generated suggestions for expanding the ``Projections'' node. (c) Query-driven expansion requesting examples of climate change causes. (d) Command-based expansion of industrial processes. (e) Chat-based query adding adaptation strategies. (f) Direct manipulation: creating a custom node and linking it to existing concepts. (g) Natural language contribution: adding a new node and defining its relationship through text commands. }
    \Description{ Seven labeled screenshots (a through g) demonstrating \name{} functionalities: (a) initial topic generation showing a hierarchical overview, (b) system-generated expansion suggestions, (c) query-driven expansion adding child nodes, (d) command-based expansion of industrial processes, (e) chat-based query adding adaptation strategies, (f) direct manipulation creating a custom node, (g) natural language contribution adding a node with a typed relationship.}
    \label{fig:use_case}
\end{figure*}

\begin{figure*}[t]
    \centering
    \includegraphics[width=1\linewidth, page=1]{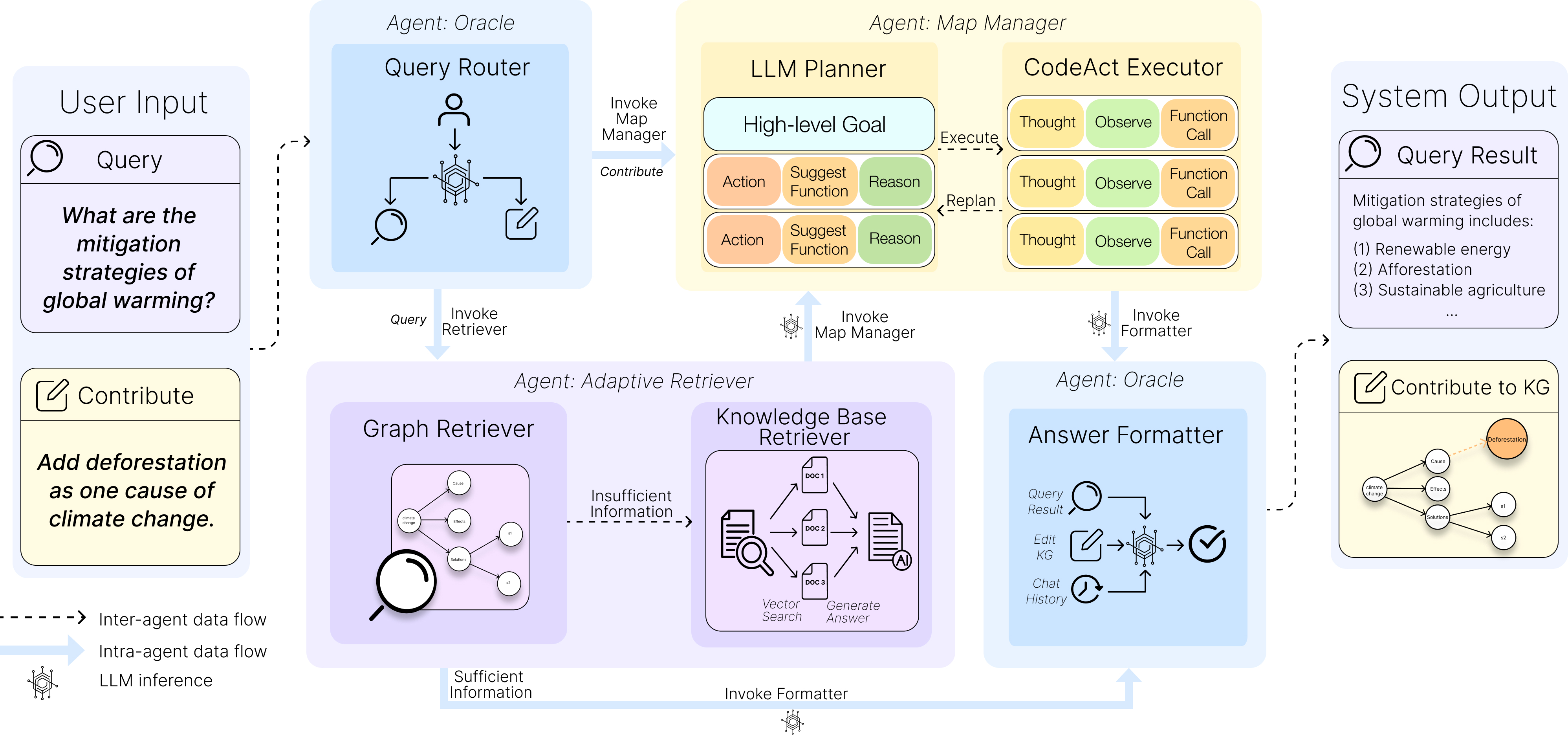}
    \vspace{-7mm}
    \caption{Overview of the \name{} multi-agent pipeline supporting bidirectional interaction. 
    User input is routed by the Oracle to either the Query Pathway (handled by the Adaptive Retriever) or the Contribution Pathway (handled by the Map Manager).
    Both pathways interact with the co-created knowledge graph, which evolves through iterative user engagement.
    The Oracle coordinates responses and maintains conversational context across interactions.
    }
    \Description{Architecture diagram of the multi-agent pipeline. User input flows to the Oracle agent, which routes queries to the Adaptive Retriever and contributions to the Map Manager. The Retriever contains a Graph Retriever and Knowledge Base Retriever with a grading step. The Map Manager contains an LLM Planner and CodeAct Executor with a replan loop. Both pathways output to the system response.}
    \label{fig:workflow}
\end{figure*}

Guided by the design goals established in Sec.~\ref{sec:design_goals}, we developed \name{} to enable human-AI collaborative knowledge construction. 
Achieving this goal requires addressing three technical challenges: first, user input is inherently ambiguous, as the same natural language statement could be a query seeking information or a contribution intended to modify the knowledge structure; second, users navigating unfamiliar topics need information at varying levels of granularity; and third, when users contribute new knowledge, the system must determine appropriate placement within the existing structure while maintaining coherence. 

To address these challenges, we developed a multi-agent architecture that coordinates specialized components: an Oracle for intent classification, an Adaptive Retriever that supports variable-granularity retrieval, and a Map Manager for coherent knowledge placement. These components operate on a shared knowledge graph that evolves through user interaction.

\subsection{System Overview}

\name{} enables users to both retrieve information from and contribute insights to an evolving knowledge graph through two complementary pathways (Fig.~\ref{fig:teaser}, Fig.~\ref{fig:workflow}).
In the \textbf{query pathway}, users pose questions to explore the knowledge base. The system searches for relevant information, synthesizes a response grounded in the uploaded documents, and optionally expands the knowledge graph with newly retrieved content. This pathway supports exploratory information seeking, allowing users to progressively deepen their understanding of unfamiliar topics (\textbf{D1}, \textbf{D2}).
In the \textbf{contribution pathway}, users add new concepts, modify existing relationships, or reorganize the knowledge structure to reflect their developing understanding. The system interprets these contributions, determines appropriate placement within the existing hierarchy, and integrates the changes while maintaining structural coherence. This pathway enables users to externalize their insights and shape the knowledge representation according to their own mental models (\textbf{D2}, \textbf{D3}).
These two pathways interact with a shared \textbf{knowledge graph} that combines the hierarchical structure of mind maps with the semantic relationships of concept maps. The graph evolves continuously as users query and contribute, capturing both the information retrieved from documents and the organizational decisions made by users. This co-created representation accumulates the user's explorations and serves as a lasting artifact that reflects their learning journey (\textbf{D1}).

To handle the complexity of supporting both pathways, \name{} employs a \textbf{multi-agent pipeline} consisting of three specialized components: the Oracle coordinates user interactions and classifies intent; the Adaptive Retriever searches the knowledge base at varying levels of granularity; and the Map Manager interprets contribution commands and executes modifications to the graph. These components collaborate to provide seamless bidirectional interaction, detailed in Sec.~\ref{sec:pipeline}.

Users engage with \name{} through two complementary interaction techniques: \textbf{direct manipulation} on the canvas (Fig.~\ref{fig:use_case}f), where users can create nodes, drag connections, and reorganize the layout; and \textbf{natural language} in the chat panel (Fig.~\ref{fig:use_case}g), where users can issue queries, request expansions, or specify modifications through conversational commands. This combination accommodates diverse interaction preferences while supporting both exploration and contribution (\textbf{D3}), detailed in Sec.~\ref{sec:user-interface}.

\subsection{User Scenario}

Before detailing the technical components, we present a user scenario to demonstrate how \name{} supports knowledge exploration and construction. 
Suppose Robert is an undergraduate student preparing for a course on climate change. 
With no prior knowledge of the subject, he uploads documents and readings shared by his professor and uses \name{} to explore the key concepts.

\textbf{Uncertainty About Initial Questions.}
At the outset, Robert feels overwhelmed by the unfamiliar content and is unsure where to begin (\textbf{C1}). 
He selects one of the suggested questions in the chat panel: ``What are the main topics covered in the documents?'' 
The system generates an initial node diagram that visually represents key topics in a hierarchical structure, with labeled edges indicating relationships between concepts. 
Each node includes a title and additional details for further exploration (Fig.~\ref{fig:use_case}a).

As Robert reviews the graph, he becomes curious about ``Projections'' but is uncertain about its subtopics. 
He clicks the suggestion button \wsicon{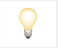} on the node toolbar, which provides four options for related topics (Fig.~\ref{fig:use_case}-b2).
Intrigued by ``climate feedback mechanisms,'' he selects it, and the system adds a new node linked to ``Projections'' with an edge indicating that climate feedback mechanisms are a component of climate projections (Fig.~\ref{fig:use_case}-b3). 
This feature allows Robert to expand the graph via system-generated suggestions, even when he lacks familiarity with the content (\textbf{C1}, \textbf{D2}).

\textbf{Knowing What to Ask.}
Robert decides to learn about the causes of climate change. 
He clicks the plus icon \wsicon{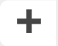} on the ``Causes'' node and types ``Give me five examples.'' 
The system adds a new node titled ``Major Contributions to Climate Change'' with the relationship ``includes,'' and five examples such as ``Deforestation,'' appear as child nodes (Fig.~\ref{fig:use_case}c).
The question and response are also recorded in the chat panel for reference (Fig.~\ref{fig:use_case}-c4).

One example, ``Industrial Processes,'' catches Robert's interest. 
He types ``expand this,'' and the system adds detailed nodes about specific industrial processes contributing to CO2 emissions (Fig.~\ref{fig:use_case}d).
Next, Robert types ``List strategies of adaptation'' into the chat, and the system responds with relevant strategies, adding them as child nodes under ``Adaptation Strategies'' (Fig.~\ref{fig:use_case}e).
By progressively exploring topics, Robert continues to expand his understanding. The hierarchical structure allows Robert to navigate complex concepts with ease, maintaining a clear overview of related information. Unlike traditional workflows that require switching between a chat interface and a separate mapping tool (\textbf{C5}), \name{} integrates exploration and visualization in a unified environment (\textbf{D1}, \textbf{D2}).

\textbf{Contributing External Knowledge.}
Robert finds video content on climate change and its business implications that he wants to integrate into his exploration. 
In conventional systems, users cannot modify the underlying knowledge base (\textbf{C3}); they can only retrieve information but not contribute their own insights.
With \name{}, Robert creates a custom node using the toolbar button \wsicon{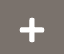} for ``financial investment in clean technologies'' and links it to ``Greenhouse Gas Emissions'' (Fig.~\ref{fig:use_case}f), demonstrating direct manipulation to add new content (\textbf{D3}).

He then wants to incorporate another insight from the video: that emissions data increasingly influences investment decisions. 
Rather than manually locating where this concept belongs in the growing graph (\textbf{C6}), Robert simply types into the chat: ``Greenhouse gas emissions are becoming important for financial investment decisions,'' followed by a detailed explanation.
The system recognizes this as a contribution rather than a query, analyzes the existing graph structure to identify ``Greenhouse Gas Emissions'' as the relevant anchor node, creates a new node capturing the concept, and establishes a relationship labeled ``is important for''—all without requiring Robert to specify commands or node names (Fig.~\ref{fig:use_case}-g).

When Robert reviews the result, he notices the system placed the new node as a child of ``Greenhouse Gas Emissions.'' He prefers it as a sibling node instead, so he types ``move this node to be at the same level as Greenhouse Gas Emissions.'' The Map Manager re-plans the graph structure and executes the modification, demonstrating how users can refine the system's intelligent placement decisions through natural conversation.

This interaction illustrates the agentic nature of the contribution pathway: users express their knowledge naturally, and the system handles intent classification, anchor identification, relationship inference, and structural placement, transforming conversational input into coherent graph modifications (\textbf{C3}, \textbf{C6}, \textbf{D2}, \textbf{D3}).

\textbf{Revisiting Previous Explorations.}
Later, Robert wants to revisit information about ``carbon sinks'' but cannot locate it in the expanded graph (\textbf{C2}, \textbf{C4}). 
He types ``What are carbon sinks?'' into the chat. 
Since this information already exists in the graph, the system responds without creating a duplicate node, helping Robert access the information without redundancy.

By the end of his session, Robert had built a well-structured representation of climate change concepts using \name{}. 
The system allowed him to explore relationships between topics while organizing information into a coherent structure that serves as a valuable reference for his course.

\subsection{Knowledge Representation for Human-AI Co-Creation} \label{sec:knowledge-rep}

We design the knowledge graph in \name{} to serve dual purposes: it should be intuitive for users to navigate and understand, while also being structured enough for the system to manipulate coherently when integrating user contributions. 
To achieve this, we integrate features from both mind maps and concept maps. This combined representation supports human comprehension through familiar hierarchical organization, while enabling AI-assisted construction through explicit semantic structure (\textbf{D1}).

From \textbf{mind maps}, the knowledge graph inherits a hierarchical, top-down approach that allows users to visualize relationships between topics clearly. 
Nodes represent concepts, and parent-child links depict hierarchical relationships, creating a natural exploration path from general topics to specific details.

From \textbf{concept maps}, the knowledge graph incorporates semantic edge labels that provide context on how concepts relate. 
For example, when exploring climate change, a user might encounter the node ``Greenhouse Gas Emissions'' connected to ``Fossil Fuel Combustion'' with an edge labeled ``is caused by,'' and to ``Carbon Pricing Policies'' with an edge labeled ``can be mitigated through.'' 
These labels allow users to grasp the nature of relationships at a glance without reading detailed content.

The graph also supports \textit{common child nodes} that connect to multiple parents, representing cross-cutting concepts relevant to several categories. 
For instance, ``Deforestation'' relates to both ``Causes of Climate Change'' (with relationship ``contributes to'') and ``Biodiversity Loss'' (with relationship ``leads to''), capturing how a single phenomenon connects to multiple broader themes without duplicating the node.

This combination enables effective co-creation: the hierarchical structure helps both users and the system navigate the graph and identify appropriate locations for new content, while semantic labels help both parties understand relationships: users can quickly comprehend the knowledge structure, and the system can leverage semantics to suggest meaningful placements for contributions.

Through iterative user interaction, the knowledge graph evolves progressively and captures the user's developing understanding (\textbf{D2}). 
Rather than presenting all information at once, the system expands the graph incrementally in response to user queries and contributions. 
Such incremental expansion aligns with exploratory information seeking, where users prefer to encounter information gradually as their familiarity grows, preventing cognitive overload while supporting continuous discovery.

\subsection{Multi-Agent Pipeline for Bidirectional Interaction} \label{sec:pipeline}



To address the challenges of intent ambiguity, granularity matching, and coherent placement, we developed a multi-agent pipeline comprising three specialized components: the Oracle for intent classification and routing, the Adaptive Retriever for variable-granularity knowledge retrieval, and the Map Manager for coherent knowledge placement (Fig.~\ref{fig:workflow}).

\textbf{Oracle: Intent Classification and Routing.}
The Oracle serves as the central coordinator that classifies user intent and routes requests to the appropriate pathway. 
It analyzes each input with a large language model to determine whether the input represents a \textit{query} (seeking information) or a \textit{contribution} (modifying the knowledge graph). 
The classification considers linguistic cues, including imperative structures and explicit commands like ``add'' or ``link'' suggest contributions, while interrogative forms suggest queries, as well as conversational context from previous interactions.
For queries, the Oracle invokes the Adaptive Retriever and formats responses for the user. 
For contributions, it directs the request to the Map Manager and reports the outcome.

\begin{figure}[tb]
    \centering
    \includegraphics[width=\linewidth, page=1]{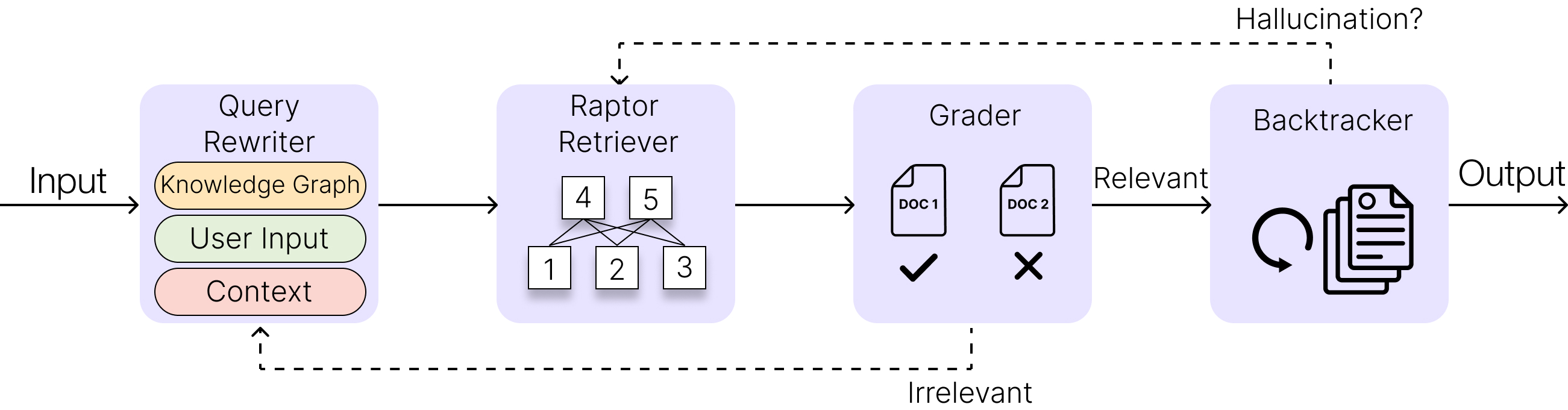}
    \vspace{-6mm}
    \caption{The Adaptive Retriever pipeline consists of four distinct stages: (i) Query Rewriter: refines the user's input to optimize it for retrieval, (ii) Raptor Retriever: retrieves relevant text chunks at various levels of granularity by recursively embedding and clustering information, (iii) Text Chunks Grader: evaluates the retrieved chunks for relevance to the user's query, filtering out less pertinent content, and (iv) Backtracker: links the generated response back to the original source, providing metadata to ensure transparency and grounding of information.
    }
    \Description{Pipeline diagram of the Adaptive Retriever showing four sequential stages: Query Rewriter taking input with knowledge graph and user context, Raptor Retriever searching hierarchical text chunks, Grader filtering relevant documents, and Backtracker linking responses to source documents with a hallucination check loop. }
    \label{fig:retriever}
\end{figure}

\textbf{Adaptive Retriever: Variable-Granularity Knowledge Retrieval.}
Users exploring unfamiliar topics need information at different levels of detail—broad overviews when orienting themselves, specific facts when investigating particular concepts. 
Standard retrieval-augmented generation (RAG) approaches retrieve a fixed number of text chunks, which proves insufficient for summative questions requiring broader context and provides no mechanism for adapting to user familiarity.

To address this, we employ a Raptor retriever~\cite{sarthi2024raptor} that enables retrieval at varying granularities. 
The retriever recursively embeds, clusters, and summarizes text chunks into a hierarchical tree structure. 
When users are new to a topic, the retriever draws from higher levels of the tree to provide broader overviews. 
As users become more familiar, it retrieves from lower levels to supply detailed, grounded facts.

The retrieval pipeline (Fig.~\ref{fig:retriever}) includes four stages: a \textit{Query Rewriter} optimizes the user's input for vector database retrieval; the \textit{Raptor Retriever} searches for relevant chunks at appropriate granularity; a \textit{Grader} evaluates retrieved chunks for relevance and filters out tangential content; and a \textit{Backtracker} links responses to source documents with metadata (title, page number) to ensure transparency and reduce hallucination risk.

\textbf{Map Manager: Coherent Knowledge Placement.}
When users contribute new knowledge, the system needs to interpret their intent and integrate the contribution coherently. 
A node about ``deforestation'' should connect to ``causes of climate change'' with an appropriate relationship label, not appear as an orphan or attach to an unrelated concept.

The Map Manager handles contribution processing through a plan-and-execute scheme with iterative error correction (Algorithm~\ref{alg:map-manager}). 
In the \textit{planning} phase, an LLM interprets the user's objective, analyzes the current graph state, and generates a sequence of function calls to achieve the goal. 
The knowledge graph is represented internally as an adjacency matrix, and the planner produces operations such as \texttt{AddNode()}, \texttt{AddEdge()}, \texttt{DeleteNode()}, or \texttt{UpdateNode()} with appropriate arguments.

In the \textit{execution} phase, a CodeAct agent~\cite{wang2024codeact} executes each planned operation sequentially. 
If an operation fails (e.g., target node not found, invalid argument format), the executor attempts self-correction by reasoning from the error message and conversation context. 
If self-correction fails after $N$ attempts (we use $N=5$), the executor reports the error back to the planner, which re-plans based on the current graph state and error information. 
This plan-execute-replan cycle continues until all operations complete successfully or the maximum recursion depth is reached.

\begin{algorithm}[htbp]
\caption{Map Manager: Plan-and-Execute with Self-Correction}
\label{alg:map-manager}
\SetKwInOut{Input}{Input}
\SetKwInOut{Output}{Output}
\SetKw{Break}{break}
\SetKw{Goto}{goto}
\SetAlgoNoEnd  
\DontPrintSemicolon  
\Input{$u$: user input; $G$: current graph state; $H$: conversation history}
\Output{$G'$: updated graph; $r$: response to user}
$depth \leftarrow 0$\;
\While{$depth < MAX\_DEPTH$}{
    $ops \leftarrow \texttt{Plan}(u, G, H)$ \tcp*[r]{e.g., [AddNode(...), AddEdge(...)]}
    \ForEach{$op \in ops$}{
        \For{$attempts \leftarrow 0$ \KwTo $N$}{
            $result \leftarrow \texttt{Execute}(op, G)$\;
            \lIf{$result$.success}{$G \leftarrow result.newState$; \Break}
            \lElse{$op \leftarrow \texttt{SelfCorrect}(op, result.error, H)$}
        }
        \lIf{$attempts = N$}{$H \leftarrow H \cup \{error\}$; $depth$++; \Goto line 3}
    }
    \Return{$G$, \texttt{FormatResponse}($ops$)}\;
}
\Return{$G$, ``Max attempts reached.''}\;
\end{algorithm}

Consider a user who inputs ``Greenhouse gas emissions are becoming important for financial investment decisions.'' The Oracle classifies this as a contribution based on its declarative structure expressing a relationship. The Map Manager's planner analyzes the graph, identifies ``Greenhouse Gas Emissions'' as the anchor node, and generates two operations: creating a new node titled ``Financial Investment Relevance'' and establishing an edge labeled ``is important for.'' The executor runs each operation; if, for instance, the edge creation fails because the relationship type is not recognized, the self-correction mechanism reasons from the error and retries with a corrected argument (e.g., changing to ``influences''). Upon successful completion, the Oracle confirms the update.

This architecture provides several advantages over single-prompt approaches: specialized components can be refined independently, error handling is more robust through the plan-execute-replan cycle, and the separation between intent classification, retrieval, and graph modification enables cleaner reasoning at each stage. 
Detailed prompts for each component are provided in Appendix~\ref{apx:prompts}.

\subsection{User Interface} \label{sec:user-interface}

The \name{} interface comprises three main components (Fig.~\ref{fig:teaser}): a file panel for document management, a canvas for knowledge graph visualization and direct manipulation, and a chat panel for natural language interaction.

\textbf{File Section.}
The file panel (Fig.~\ref{fig:teaser}a) allows users to upload and organize documents for their knowledge base. 
Users can create folders by clicking the folder icon \wsicon{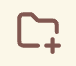} and dragging files into the desired location. 
The system automatically processes uploaded documents and stores them in the vector database for retrieval.

\textbf{Chat Section.}
The chat panel (Fig.~\ref{fig:teaser}c) supports natural language queries and contributions. 
At the start of a session, the system provides suggested questions to help users begin exploration. 
Users can type queries to retrieve information (e.g., ``What are the causes of climate change?'') or issue commands to modify the knowledge graph (e.g., ``expand this,'' ``delete,'' ``link X to Y''). 
Each response appears in the chat along with source references, and corresponding changes are reflected on the canvas immediately.

\textbf{Widget Bar.}
The widget bar at the top of the interface provides quick access to key functions. 
The ``File Manager'' \wsicon{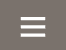} and ``Chat with AI'' \wsicon{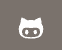} buttons toggle the side panels. 
The ``Create Node'' button \wsicon{create_custom_node} allows users to add custom nodes for notes or external knowledge. 
The ``Group Nodes'' button \wsicon{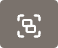} enables users to cluster related nodes; once grouped, users can ungroup them or remove individual nodes with the scissor icon \wsicon{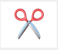}. 
The ``Rearrange Canvas'' button \wsicon{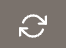} automatically reorganizes the layout when the graph becomes cluttered.

\textbf{Canvas Section.}
The canvas (Fig.~\ref{fig:teaser}b) displays the knowledge graph and supports direct manipulation for both exploration and contribution. 
When a user selects a node, a toolbar appears with four functions: the plus icon \wsicon{plus} opens an input box for queries focused on that node; the delete icon \wsicon{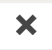} removes the node; the suggestion icon \wsicon{light_bulb} provides expansion recommendations; and the star icon \wsicon{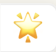} highlights important nodes in yellow.

The canvas also features \textit{semantic zoom} to help users navigate large graphs (\textbf{D1}). 
At a zoomed-out level, only node titles are visible, which provides a high-level structural overview. 
As users zoom in, detailed content within each node becomes readable, allowing fluid transitions between broad orientation and focused exploration.
\revision{The graph uses a horizontal hierarchical layout based on breadth-first traversal, with root nodes on the left and child nodes extending rightward.
Users can drag nodes to override computed positions; the ``Rearrange Canvas'' button recomputes the full layout from the current graph topology when the arrangement becomes cluttered.}

\remark{change log: Added layout management description per R2's comments}

\subsection{Implementation Details} \label{sec:implementation_details}
\name{} consists of a React.js\footnote{https://react.dev} frontend and a Flask backend. 
For language model inference, we use GPT-4o\footnote{https://platform.openai.com/docs/models/gpt-4o} for planning tasks in the Oracle and Map Manager, and GPT-4o-mini for simpler tasks to reduce latency. 
The backend workflow is implemented using LangGraph~\cite{langgraph}, which abstracts the multi-agent coordination as a state machine. 
For the Adaptive Retriever, we use LangChain's ChromaDB wrapper\footnote{https://python.langchain.com/docs/integrations/vectorstores/chroma/} to construct the vector store and OpenAI's text-embedding-3-small\footnote{https://platform.openai.com/docs/models/text-embedding-3-small} for computing embeddings. 
The knowledge graph data structure was implemented from scratch to support the hierarchical and semantic features described in Section~\ref{sec:knowledge-rep}. 
For the user study, we deployed \name{} on Microsoft Azure.
\section{User Study}

We conducted a controlled user study to evaluate \name{}'s effectiveness in supporting users' knowledge management and information seeking by comparing against a baseline.

\subsection{Participants}
We recruited 12 participants (6 female, 6 male, aged 18-35), including 5 Master's students, 4 PhD students, and 3 undergraduates. They had diverse backgrounds spanning computer science, neuroscience, data science, and human-computer interaction. All participants had prior experience with generative AI tools such as ChatGPT and Gemini, and most reported using these tools frequently. 
Additionally, the majority had experience using mind maps for organizing or learning information. 
The experiments were performed on a laptop computer equipped with a mouse. 
Each participant received \$30 for their participation.

\subsection{Design}
We employed a within-subject approach to compare two systems, \name{} and a baseline, by curating two study tasks over two different datasets, using a counterbalanced design.

\textbf{Datasets.}
The tasks required participants to create a slide deck by using the systems to explore and understand the concepts from the datasets. 
We selected datasets covering two distinct domains: climate change and AI ethics. 
For each topic, we compiled a collection of related documents sourced from Wikipedia. 
The climate change dataset consisted of six documents, while the AI ethics dataset included five documents. 
Each document contained between 2,000 and 4,000 words. 
To ensure compatibility with our text-only retrieval system, we removed all figures from the articles, as our current RAG implementation only supports textual content.

\textbf{Baseline.} 
Our baseline was a retrieval-only system with graph visualization capabilities (see Appendix~\ref{apx:baseline}). 
The baseline consisted of two components: the same Adaptive Raptor Retriever used in \name{} for querying the document corpus, and a visualization interface that automatically converts LLM responses into static node-link diagrams. 
Participants first selected a dataset, then queried it through the retriever interface.
After receiving answers, they could transfer the results to the visualization interface, which generated corresponding concept maps. 
Participants could then ask additional questions and examine the visualization.

We chose this baseline to isolate the effect of the contribution pathway, as both systems support retrieval and visualization, but only \name{} allows users to contribute to and modify the knowledge structure. 
This design choice enables us to evaluate whether bidirectional interaction (querying \textit{and} contributing) improves knowledge organization compared to retrieval-only interaction. 
We acknowledge that comparing systems with different capabilities introduces potential confounds; however, our goal was to evaluate whether the contribution pathway improves knowledge organization, which required a baseline without this capability.

\subsection{Procedure}
The study lasted approximately two hours for each participant, consisting of the following steps.

\textbf{Introduction and Consent (15 minutes).}
Initially, we provided a brief introduction to the study background, goals, and the two systems, and administered a demographics pre-questionnaire.

\textbf{System Tutorials (10 minutes each).}
Before using each system, we provided an instructional video that introduced its key features, giving participants a basic understanding of the interface and functionality.
Participants were then given time to interact with the system they would use for the upcoming task. 
During this time, we guided them through the features, showing them how to explore and build knowledge graphs. 
A test dataset on the topic of operating systems was used during this familiarization phase.

\textbf{Study Tasks (25 minutes each).}
A Latin square counterbalancing method was used to alternate between systems and datasets across participants. 
Participants were told: ``Imagine you have been invited to give a presentation on the selected topics. 
You will use Google Slides to create slide decks on selected topics. 
Each slide deck should be created by exploring and understanding the content provided in the documents using the given system. 
You do not need to add images or any other visual decorations. 
You will have 20 minutes to work with each system.'' 
The participants' interactions with the systems, as well as their process of building the slide decks, were recorded.

\textbf{Post-questionnaire and Interview (20 minutes).} 
After each task, participants completed a post-questionnaire to evaluate the usefulness and effectiveness of the systems, along with specific feature ratings for \name{}. 
At the end of the study, we conducted a 10-15 minute semi-structured interview where participants compared the two systems based on their experiences. 
They also provided insights on specific features of \name{}.

\section{User Study Results}

\begin{figure*}[tb!]
    \centering
    \includegraphics[width=\linewidth, page=1]{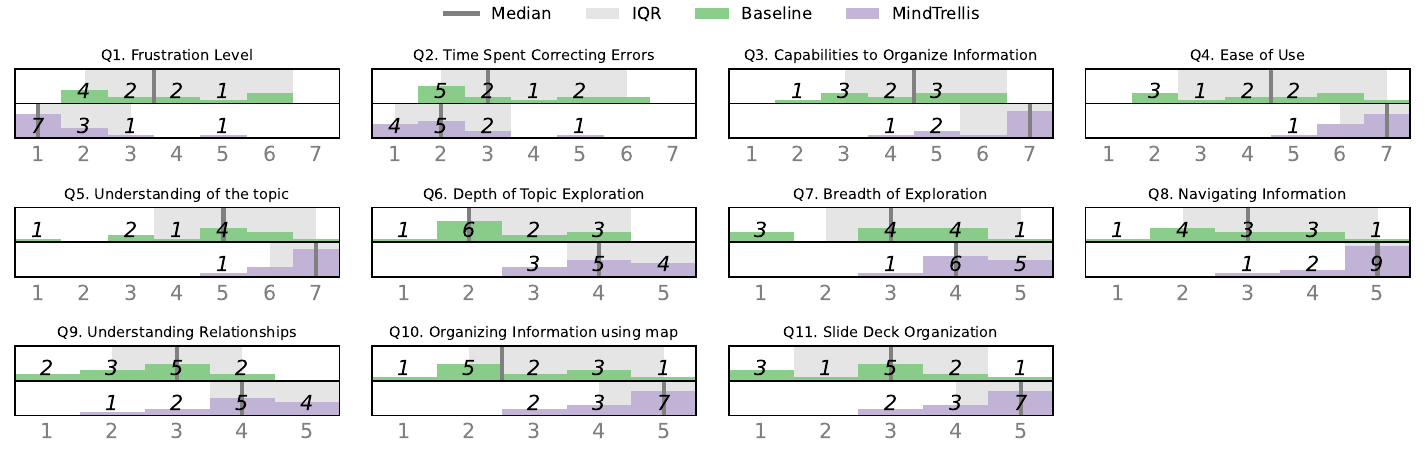}
    \vspace{-8mm}
    \caption{Participants' ratings on usability (UMUX), task support, and depth and breadth of information exploration. Usability measures (Q1--5) were rated on a 7-point Likert scale, assessing whether \name{} met their requirements, ease of use, and frustration experienced. Task-related questions (Q6--11) were rated on a 5-point Likert scale, assessing how well the system supported understanding of the topic, the detail and breadth of information provided, and the system's capability in organizing slide deck content. For Q1 and Q2, lower scores are better; for the remaining questions, higher scores are better.}
    \Description{Bar chart comparing participant Likert ratings between Baseline (gray) and \name{} (green) across eleven questions Q1 through Q11, covering frustration, error correction, organization capability, ease of use, topic understanding, depth and breadth of exploration, relationship understanding, map organization, and slide deck organization. \name{} receives higher ratings on all positively-framed questions.}
    \label{fig:questionnaire_1}
\end{figure*}

\begin{figure}[tb!]
    \centering
    \includegraphics[width=\linewidth, page=1]{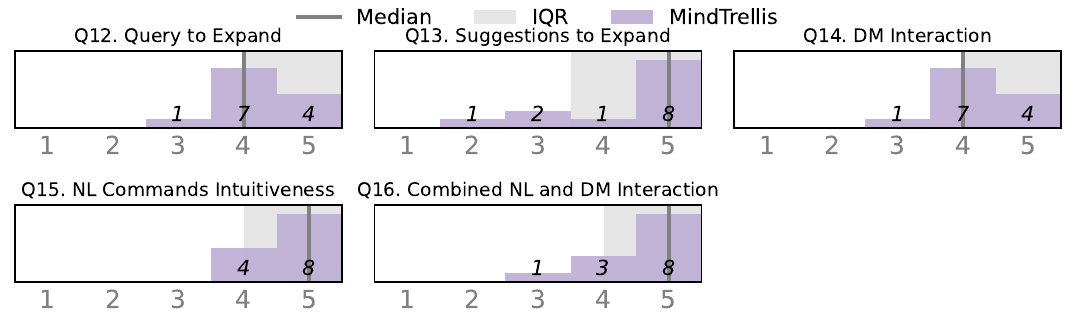}
    \vspace{-6mm}
    \caption{Participants' ratings on \name{}'s effectiveness of knowledge expansion, system suggestions, and user interaction methods (direct manipulation and natural language commands). Ratings were collected on a 5-point Likert scale.}
    \Description{Bar chart showing \name{}-only Likert ratings across five questions Q12 through Q16, covering query expansion, suggestion expansion, direct manipulation, natural language intuitiveness, and combined interaction. Most ratings cluster at 4 and 5 on the 5-point scale.}
    \label{fig:questionnaire_2}
\end{figure}

We analyzed the results using a mixed-method approach that included system usage logs, post-questionnaires, and semi-structured interviews.
We compared \name{} with the baseline across usability, effectiveness of the generated knowledge graphs, and support for knowledge exploration using the Wilcoxon signed-rank test (Fig.~\ref{fig:questionnaire_1}: Q1--11 and Fig.~\ref{fig:questionnaire_2}: Q12--16).
Below, we report findings organized by our three design goals (D1--D3).

\subsection{Co-Created Knowledge Graph Supports Exploration (D1)}

Participants reported that the co-created knowledge graph enhanced their exploration by providing a structure they could shape together with AI assistance.
The co-created graph received higher ratings for organizing information (Q3: $Mdn_M = 7.0 > Mdn_B = 4.5$, $p < 0.01$, $r = 0.883$), ease of use (Q4: $Mdn_M = 7.0 > Mdn_B = 4.5$, $p  < 0.01 $, $r = 0.883$), and supporting topic understanding (Q5: $Mdn_M = 7.0 > Mdn_B = 5.0$, $p < 0.01$, $r = 0.883$).

The co-created graph supported exploration in three main ways.
First, participants found that being able to shape the visual organization made material ``easier to comprehend'' (P4) and allowed them to ``organize [the] knowledge graph fully before starting on slides'' (P1).
P12 appreciated how the system ``allows me to control the flow of the graph that follows my own mindset,'' emphasizing the collaborative nature of the representation.
Second, the structure enabled in-depth exploration while maintaining coherence.
P8 could focus on ``higher-level ideas'' and P9 could ``expand on key aspects of a topic in a consistent manner'' (Q9: $Mdn_M = 4.0 > Mdn_B = 3.0$, $p <0.05$, $r = 0.883$).
Third, detailed content with titles, explanations, and semantic edge labels simplified comprehension.
P2 noted that seeing ``detailed relationships between nodes [on edges]'' enabled deeper understanding, and P5 found that ``relevant examples'' made complex topics ``simpler and more straightforward.''

In contrast, the baseline generated graphs automatically without user input, producing structures that participants found difficult to work with.
P2 described the baseline as offering only ``vague relationships,'' and P9 found the information ``scattered,'' making it ``challenging to maintain a structured overview.''
P1 wanted ``the ability to organize'' when using the baseline, and P5 noted that ``the space is really limited, and I couldn't freely zoom in or zoom out to see the overview.''
Unlike the baseline's fixed, auto-generated structure, the co-created graph introduces information progressively as users explore, reducing initial overwhelm (Q1: $Mdn_M = 1.0 < Mdn_B = 3.5$, $p < 0.01$, $r = 0.815$).
Participants also reported that \name{} better supported creating ``in-depth slides and connections across topics'' (P3) (Q11: $Mdn_M = 5.0 > Mdn_B = 3.0$, $p < 0.01 $, $r = 0.883$).
These observations confirm the success of D1 in supporting exploration through a co-created, visually structured knowledge graph.

\subsection{Bidirectional Interaction Enables Cumulative Knowledge Building (D2)}

A key capability of \name{} is bidirectional interaction: users can both query from and contribute to the evolving knowledge structure.
This section reports how participants engaged with both pathways and how the iterative cycle between them supported cumulative knowledge building.

\subsubsection{Query Pathway: Retrieving and Exploring Information}

Participants used the query pathway to retrieve information from their uploaded documents in four ways.
First, they asked questions to retrieve information directly.
P7 noted that ``I could just type my question and get relevant information without searching through all the documents myself.''
Second, they issued commands to expand specific nodes.
P2 ``liked the option to extend nodes using natural language'' because it allowed them to ``expand on ideas directly and get more specific examples'' (Q12: $Mdn = 4.0$, $IQR = 1.0$).
Third, they requested context about specific nodes without losing the overall structure.
P8 shared that ``it allowed me to engage with specific nodes, which made learning about each topic more in-depth as I could explore related concepts seamlessly'' (Q15: $Mdn = 5.0$, $IQR = 1.0$).
Unlike the baseline, which changed the graph structure upon additional operations, \name{} preserved context during querying (Q2: $Mdn_M = 2.0 < Mdn_B = 3.0$, $p < 0.05$, $r = 0.758$).
Fourth, they used the suggestion feature for targeted inspiration.
P11 appreciated that ``it is helpful when I'm new to a topic and don't know what should be included next. It gives me a direction to expand'' (Q13: $Mdn = 5.0$, $IQR = 1.25$).
P10 could ``quickly explore and expand the knowledge graph without having to manually add nodes,'' and P6 noted efficiency gains: ``it allowed me to see suggested connections and explore those connections quickly.''

\subsubsection{Contribute Pathway: Adding and Refining Knowledge}

Beyond retrieval, participants actively contributed their own knowledge in two main ways.

First, when they had relevant domain expertise or prior knowledge, participants added new concepts that extended beyond the source documents. 
P12, a participant with a finance background, discovered connections between climate change and financial investment. 
They supplemented the knowledge base with their domain expertise: ``I'm happy 
it made my slide more concrete.'' Similarly, P3 added concepts from prior coursework: ``I added some nodes based on what I already knew, which helped me connect the new material to my existing understanding.''

Second, when the system's organization did not match their mental models, participants reshaped the structure. 
For fine-grained adjustments, participants used direct manipulation to drag nodes and visually group concepts. 
P11 emphasized: ``It gives me control over how I explore and arrange the information, which is essential for structuring my thoughts'' (Q14). 
However, P9 noted that ``expanding groups together would be a great addition,'' suggesting that batch operations on grouped nodes could further enhance direct manipulation.
For bulk modifications---such as reorganizing entire sections or modifying multiple edge labels---natural language proved more efficient. 
P10 noted that ``typing a command was faster than dragging things around.''


\subsubsection{Cumulative Building Through Iterative Interaction}

Participants fluidly moved between querying and contributing, and this iterative cycle enabled cumulative knowledge building.
Rather than completing all queries before contributing or vice versa, participants alternated both pathways throughout their sessions.

Two dominant patterns emerged.
In the first pattern, participants started with queries and transitioned to contribution when they identified gaps.
P4 described this cycle: ``I can keep asking questions, and they can keep inspiring me with new topics. It is easier to explore and expand on topics cohesively.''
For example, P12, a participant with a finance background, was exploring climate change documents to prepare a presentation.
While reviewing content on greenhouse gas emissions, they recognized connections to their prior expertise---specifically, how emissions data relates to financial investment decisions such as ESG (environmental, social, governance) investing.
Although the source documents did not cover this angle, they supplemented the knowledge base with their domain knowledge: ``I'm happy it made my slide more concrete.''
They then queried further to expand those newly added concepts, building out the financial investment branch of the graph.

In the second pattern, participants contributed first and used queries to expand from their additions.
P3 noted: ``I added some nodes based on what I already knew, then asked the system to give me more details about those areas.''
This contribute-then-query approach allowed participants to anchor exploration in their existing understanding, using the system to fill in details around concepts they introduced.

The suggestion feature often triggered transitions between pathways.
P5 observed: ``Combining features like the suggestion tool and chat allowed me to explore the material in more depth.''
When suggestions surfaced unexpected connections, participants would contribute additional context from their knowledge, then query to explore those connections further.

Individual differences emerged in how participants balanced the two pathways.
Some primarily used queries, relying on chat and suggestions to drive exploration.
Others were heavily engaged with contribution, creating custom nodes and reorganizing extensively.
P3 noted: ``I spent a lot of time rearranging nodes and adding my own ideas because I wanted the graph to reflect how I think about the topic.''
This variance suggests that bidirectional interaction accommodates diverse knowledge-building styles.
These findings confirm D2's goal of enabling cumulative knowledge building through bidirectional interaction.

\subsection{Flexible Interaction with Transparent Provenance (D3)}

Participants alternated between natural language and direct manipulation, leveraging each mode's strengths for different task demands.
Transparent provenance enabled verification of retrieved information when needed.

A common pattern involved using natural language to add content, then switching to direct manipulation to adjust positioning.
P9 described: ``I would ask the system to add information, then drag the nodes around to organize them the way I wanted.''
Participants appreciated having both options available, choosing natural language for efficiency when precise placement was not critical, and direct manipulation when layout mattered (Q14: $Mdn = 4.0$, $IQR = 1.0$; Q15: $Mdn = 5.0$, $IQR = 1.0$; Q16: $Mdn = 5.0$, $IQR = 1.0$).

Participants also valued being able to trace retrieved information back to source documents.
P8 mentioned that ``knowing the answers came from my own documents made me more confident in the information,'' and P11 appreciated being able to ``click on the citation and see exactly where the information came from.''
P3 added that ``when I wasn't sure about something, I could always go back to the original document to check.''
However, P4 noted that ``sometimes the system didn't quite understand what I was asking for, and I had to rephrase my question,'' suggesting opportunities for improving query interpretation.

These findings confirm D3's support for flexible interaction through complementary modes, with transparent provenance enabling verification and trust in the co-created knowledge graph.

\section{Pipeline Evaluation} \label{sec:pipeline_eval}

In addition to evaluating the user experience, we validated if the multi-agent pipeline reliably supports bidirectional interaction. 
We evaluated each pipeline component in isolation using correctly-routed inputs, then the end-to-end task success across all inputs to measure user-facing reliability. Two researchers independently evaluated all 152 logged interactions from the user study sessions; inter-rater agreement was substantial (Cohen's $\kappa$ = 0.81), with disagreements resolved through discussion. All LLM-based classification in the pipeline uses GPT-4o.

\subsection{Oracle Evaluation: Intent Classification}

The Oracle classifies each user input into one of three intent categories: \textit{query} for information retrieval from the knowledge base, \textit{contribute} for modification to the knowledge graph, or \textit{expansion} for elaboration on existing nodes. Correct classification is essential because it determines which downstream component, either the Retriever or Map Manager, processes the input.

\textbf{Data and Method.} We assessed all 152 user inputs logged during the user study. Two researchers independently labeled each input, achieving Cohen's $\kappa$ = 0.84. We then compared the Oracle's classifications against these ground truth labels.

\textbf{Results.} The Oracle achieved 91.4\% overall accuracy. Table~\ref{tab:oracle} shows precision, recall, and F1 scores for each intent category. The primary source of error was confusion between \textit{query} and \textit{expansion} intents (9 cases), which is expected given their semantic overlap: both involve information retrieval, differing primarily in whether new nodes should be generated. Misclassification between \textit{query}/\textit{expansion} and \textit{contribute} was rare (4 cases). The Oracle reliably distinguishes retrieval requests from edit commands.

\begin{table}[tb]
\centering
\caption{Oracle Intent Classification Results}
\Description{Classification results for the Oracle showing precision, recall, and F1 scores for three intent categories (Query, Contribute, Expansion) with overall accuracy of 91.4\%.}
\label{tab:oracle}
\begin{tabular}{lcccc}
\toprule
\textbf{Intent Class} & \textbf{Precision} & \textbf{Recall} & \textbf{F1} & \textbf{N} \\
\midrule
Query & 0.89 & 0.92 & 0.90 & 58 \\
Contribute & 0.94 & 0.91 & 0.92 & 47 \\
Expansion & 0.91 & 0.87 & 0.89 & 47 \\
\midrule
\textbf{Overall Accuracy} & --- & --- & \textbf{91.4\%} & \textbf{152} \\
\bottomrule
\end{tabular}
\end{table}

\subsection{Map Manager Evaluation: Edit Execution}

The Map Manager executes contribution commands by parsing user intent, planning the required operations, and modifying the knowledge graph accordingly. We evaluate whether the Map Manager correctly executes these commands across multiple dimensions.

\textbf{Data and Method.} Of the 47 logged contribution commands, 43 were correctly classified by the Oracle. We examined these correctly-routed commands to isolate Map Manager performance from Oracle errors. Two researchers independently rated each command on four dimensions. 
1) \textit{Execution Success} captures whether the system completed without error. 
2) \textit{Node Correctness} rates whether the created or modified node contains correct content on a 3-point scale\footnote{Score scale: 3 indicates correct content, 2 indicates minor issues such as truncation, and 1 indicates wrong content.}. 
3) \textit{Placement Correctness} rates whether the node appears in the correct location in the hierarchy, also on a 3-point scale\footnote{Score scale: 3 indicates correct parent, 2 indicates reasonable but suboptimal location, and 1 indicates wrong location.}. 
4) \textit{Relationship Correctness} rates whether the edge label accurately describes the relationship on a 3-point scale\footnote{Score scale: 3 indicates appropriate, 2 indicates acceptable but imprecise, and 1 indicates wrong or misleading.}. 
Scores were averaged across annotators, with disagreements on execution success resolved through discussion. 
We also report the \textit{fully correct rate}: the percentage of commands where execution succeeded and all three correctness dimensions received a score of 3.

\textbf{Results.} Table~\ref{tab:components} (top section) presents the Map Manager evaluation results. The system achieved 93.0\% execution success rate; commands rarely caused system errors. Among successfully executed commands, node correctness was high ($M$ = 2.79/3), while placement and relationship correctness showed more variation ($M$ = 2.58/3 and $M$ = 2.63/3, respectively). The fully correct rate of 78.1\% indicates that more than three-quarters of user contributions were integrated exactly as intended. The most common placement errors occurred when users provided ambiguous parent references, such as ``add this under climate change'' when multiple nodes contained that phrase.

\begin{table}[tb]
\centering
\caption{Map Manager, Retriever, and End-to-End Evaluation Results}
\Description{Evaluation results for three pipeline components: Map Manager metrics (execution success, node/placement/relationship correctness), Retriever answer quality comparison (naive RAG vs. pipeline), and end-to-end success rates for query, contribution, and overall tasks.}
\label{tab:components}
\begin{tabular}{llc}
\toprule
\textbf{Component} & \textbf{Metric} & \textbf{Value} \\
\midrule
\textbf{Map Manager} & Execution Success Rate & 93.0\% \\
& Node Correctness (mean) & 2.79 / 3 \\
& Placement Correctness (mean) & 2.58 / 3 \\
& Relationship Correctness (mean) & 2.63 / 3 \\
& Fully Correct Rate & 78.1\% \\
& N & 43 \\
\midrule
\textbf{Retriever} & Answer Quality (Naive RAG) & 2.34 / 3 \\
& Answer Quality (Our Pipeline) & 2.81 / 3 \\
& Correct Rate (Naive RAG) & 58.8\% \\
& Correct Rate (Our Pipeline) & 82.5\% \\
& N & 97 \\
\midrule
\textbf{End-to-End} & Query / Expansion Success & 85.7\% \\
& Contribute Success & 80.9\% \\
& Overall Success & 83.6\% \\
& N & 152 \\
\bottomrule
\end{tabular}
\end{table}

\subsection{Adaptive Retriever Evaluation: Retrieval Quality}

\revision{To validate that the retrieval component provides sufficient answer quality for reliable co-creation, we compared our pipeline against a RAG-only baseline on the same logged queries.}

\revision{\textbf{Data and Method.} Of the 105 logged queries and expansion requests, 97 were correctly classified by the Oracle.
We re-ran these queries through both our pipeline (Raptor hierarchical retrieval with relevance grading) and a naive RAG baseline (standard chunking with vector similarity retrieval).
Two researchers rated each response on a 3-point scale\footnote{Score scale: 3 indicates correct (fully answers the query), 2 indicates partially correct (addresses the query but incomplete or contains minor errors), and 1 indicates incorrect (fails to answer or provides wrong information).} (Cohen's $\kappa$ = 0.76).}

\revision{\textbf{Results.} Our pipeline achieved higher answer quality than the naive baseline (2.81/3 vs.\ 2.34/3; correct rate 82.5\% vs.\ 58.8\%; Table~\ref{tab:components}, middle section), confirming that the hierarchical retrieval design provides sufficient accuracy for the downstream knowledge construction task.}

\remark{change log: Condensed two full paragraphs to shorter passages; reframed from ``our pipeline outperforms naive RAG'' to validation that retrieval quality meets the threshold for co-creation, per R2's comment that RAPTOR is not a contribution of this paper.}

\subsection{End-to-End Evaluation: Task Success}

The component evaluations above use correctly-routed inputs to isolate each component's performance. We now evaluate end-to-end task success across all inputs to measure user-facing reliability.

\textbf{Data and Method.} We evaluated all 152 user inputs without filtering. Two researchers independently judged whether each task succeeded. For query and expansion tasks, success required that the Oracle correctly classified the input and the generated response was rated at least 2 (partially correct or better). For contribute tasks, success required that the Oracle correctly classified the input, execution succeeded, and placement correctness was at least 2. For failed tasks, researchers attributed the failure to the responsible pipeline component: Oracle misclassification, Map Manager error, or Retriever/response error.

\textbf{Results.} The pipeline achieved 83.6\% overall success rate, with query/expansion tasks at 85.7\% and contribute tasks at 80.9\% (Table~\ref{tab:components}, bottom section). Among the 25 failed tasks, Oracle misclassification accounted for 52.0\% of failures (13 cases), Retriever/response errors for 36.0\% (9 cases), and Map Manager errors for 12.0\% (3 cases). The relatively higher proportion of Oracle and Retriever errors suggests that intent classification and answer generation are the primary areas for future improvement, while edit execution is comparatively robust.

\textbf{Summary.} These technical results establish that the multi-agent pipeline achieves reliable bidirectional interaction. The Oracle's 91.4\% accuracy ensured that user intents were correctly routed to the appropriate pipeline branch. The Map Manager's 78.1\% fully-correct rate meant that user contributions were reliably integrated into the knowledge structure. Our retrieval pipeline outperformed naive RAG on answer quality (2.81 vs.\ 2.34); the hierarchical design enables effective retrieval across varying levels of granularity. The overall 83.6\% end-to-end success rate confirms that participants experienced the intended interaction paradigm.
\section{Discussion}
\revision{Our study compared two designs for knowledge construction: a retrieval-only baseline where users queried documents and received system-generated graph visualizations, and \name{}, where users could also shape the evolving knowledge structure through contributions and reorganization.}
The significant differences across knowledge organization (Q3), depth of exploration (Q6), and slide deck organization (Q11) suggest that allowing users to co-construct the knowledge structure changes how they engage with complex information.
Retrieval-only systems such as Sensecape~\cite{suh_sensecape_2023} and Graphologue~\cite{graphologue} generate visualizations for users to navigate, but the structure remains static.
\name{} participants actively shaped the representation to reflect their evolving understanding.
 
Participants' preference for progressive expansion aligns with cognitive load research.
Paas~\cite{paas1992training} found that presenting information gradually helps learners process it more effectively in working memory.
Mayer and Moreno~\cite{mayer2003nine} identify segmentation as a key principle for reducing extraneous cognitive load.
The baseline's simultaneous presentation of all nodes created what P10 described as an ``overwhelming experience.''
In contrast, \name{}'s incremental expansion allowed participants to ``focus on the information they needed.''
P11 similarly noted that the system ``expands the flowchart directly from a parent node, which makes it easier to see the connections.''
 
The value participants placed on contributing to the knowledge structure resonates with research on external cognition and constructive learning.
Kirsh~\cite{kirsh2017thinking} argues that creating external representations amplifies cognition by offloading memory, making relationships explicit, and enabling iterative refinement.
P1's approach illustrates this pattern: they used \name{} to ``organize my mind map fully before starting on the slides.''
The act of structuring knowledge externally supported their subsequent task performance.
Educational research quantifies this benefit: Schroeder et al.'s meta-analysis~\cite{schroeder2018meta} found that students who actively constructed concept maps showed significantly higher learning gains ($g = 0.72$) compared to those who passively studied pre-made maps ($g = 0.43$).
These findings further reinforce that contribution, not just retrieval, deepens engagement.
 
\remark{change log (opening paragraphs): Reframed the first paragraph from ``isolated the effect of the contribution pathway'' to comparing two designs for knowledge construction, consistent with the revised Introduction.}

\revision{
\subsection{Situating Findings Relative to Existing Systems}
 
Commercial tools such as NotebookLM and Notion AI already support forms of knowledge-level bidirectionality---users can both query and contribute to underlying knowledge stores.
Our study results speak to what additional design choices improve the knowledge construction experience.
The significant differences on Q3 (knowledge organization) and Q6 (depth of exploration) emerged from a comparison where the key differentiator was the visual knowledge graph as a shared artifact that users could directly reshape.
Participants valued seeing the structure they were building (P8: ``allowed me to organize my thoughts as it focused on higher-level ideas'') and controlling its organization (P12: ``allows me to control the flow of the graph that follows my own mindset'').
In NotebookLM, notes and retrieval results exist in separate panes; in Notion AI, the database structure is not visualized as a semantic graph.\footnote{Based on publicly available versions as of early 2026}
Our findings suggest that contribution pathways are more effective when contributions are immediately visible within the evolving structure and spatially integrated with retrieved content, rather than residing in a separate view.
 
Beyond the question of where contributions become visible, our study has implications for the technical challenges that arise when users interact with the knowledge structure through natural language.
Systems that achieve bidirectional interaction through representational synchronization~\cite{hempel2019sketch, ye2020penrose, wu2020b2, cascaval2022differentiable} partition interaction across structurally explicit boundaries, so the mapping between user action and system response is largely deterministic and the system seldom needs to infer what the user intends.
When users both query from and contribute to a shared knowledge structure through natural language, however, the same input channel carries retrieval requests, contribution commands, and ambiguous mixtures of both.
Our pipeline evaluation reveals that intent disambiguation is the primary failure mode in this setting: Oracle misclassification accounted for 52\% of end-to-end failures (13 of 25 failed tasks).
As systems move from synchronizing two views of a single artifact to enabling users and AI to co-construct an evolving knowledge structure through natural language, the disambiguation challenge becomes a central design problem absent from the representational synchronization paradigm.
 
Our study also reveals how users balance exploration and construction when both are available within a single system.
Luminate~\cite{suh2024luminate} structures the design space of LLM outputs to support divergent exploration, and D\"{u}ck et al.~\cite{duck2025needles} support claim retrieval through multiple exploration pathways.
Our participants exhibited both exploratory and constructive patterns: some explored broadly before contributing (P1, P4, P5, P10, P12), while others anchored exploration in their own contributions from the start (P3, P6, P8, P11).
The coexistence of these patterns suggests that exploration and construction are not sequential phases but interleaved activities that reinforce each other.
Luminate's approach of surfacing diverse response dimensions could complement a system like \name{}'s persistent evolving structure---future systems might integrate structured exploration of LLM outputs as a way to seed or enrich a knowledge graph before and during user contribution.
}
 
\remark{change log (Section 8.1): New subsection to extend the discussion with cross-system comparison and implications. Three paragraphs compare findings with: (1)~commercial tools (NotebookLM, Notion AI) on visual KG as shared artifact, (2)~representational synchronization systems on intent disambiguation cost, (3)~exploration-focused tools (Luminate, D\"{u}ck et al.) on interleaved exploration and construction patterns.}

\subsection{Design Implications}
 
We identify three design implications for systems that support human-AI knowledge construction.

\textbf{First, natural-language co-creation systems need explicit mechanisms for intent disambiguation.}
\revision{In systems that synchronize two views of a single artifact through representational synchronization~\cite{hempel2019sketch, ye2020penrose, wu2020b2, cascaval2022differentiable}, the editing modality is structurally partitioned and intent interpretation is largely unnecessary.
When users instead interact with a shared knowledge structure through natural language, the same input channel carries queries, contribution commands, and ambiguous mixtures of both.}
P4 noted that ``sometimes the system didn't quite understand what I was asking for, and I had to rephrase my question.''
The 52\% share of Oracle errors among end-to-end failures confirms that disambiguation is the primary technical bottleneck for this class of system.
Deng et al.~\cite{deng2025interactcomp} observe that most language agents ``lack interactive mechanisms; when faced with ambiguity, agents confidently commit to an assumed query, leading to incorrect answers.''
Future systems should provide explicit feedback about how input was interpreted and allow easy correction when misclassification occurs.
\revision{During our study, participants who encountered misinterpretations typically discovered the mismatch only after the graph had already been modified, requiring manual correction (P4, P9).
In systems where the knowledge structure and the input interface coexist as separate representations, such as \name{}'s graph canvas and chat panel, disambiguation feedback could span both: the chat panel could present the system's interpretation and request clarification, while the graph canvas could preview the intended structural change for the user to confirm or reject before execution.}
\revision{
More generally, systems accepting unconstrained natural language input into a shared knowledge structure are likely to face similar challenges~\cite{shahriari2025nl, deng2025interactcomp}, and the design of the input interpretation layer deserves as much attention as the knowledge representation itself.
}
 
\textbf{Second, co-creation systems should support flexible information pacing and expansion granularity.}
Our findings confirm that progressive disclosure reduces cognitive load, but the more consequential design question is what granularity of expansion to offer in a user-editable structure.
P8 described how \name{} ``allowed me to organize my thoughts as it focused on higher-level ideas,'' suggesting that initial displays should emphasize conceptual structure over detail.
However, P9 also noted that ``expanding groups together would be a great addition,'' indicating that users sometimes need to reveal related content simultaneously rather than node-by-node.
\revision{Sensecape~\cite{suh_sensecape_2023} addresses multilevel abstraction, but within a read-only structure. In an editable co-created graph, expansion granularity interacts with the user's ongoing reorganization---expanding a cluster may conflict with manual rearrangements the user has already made.}
Future systems should support flexible expansion at multiple levels---individual nodes, related clusters, or entire subtrees---and allow users to control the pacing of information revelation as the structure grows.
 
\revision{\textbf{Third, AI-generated organizational structure should be treated as provisional and user-adjustable at every level.}}
\revision{Previous systems for LLM-augmented knowledge exploration~\cite{graphologue, suh_sensecape_2023, suh2024luminate} generate knowledge structures that users cannot reorganize---the AI's organizational decisions are final.
Our study shows that participants actively reshaped the AI's structure across both interaction modalities.
P7 and P10 used natural language to reorganize portions of the graph; P10 noted that ``typing a command was faster than dragging things around.''
P9 used direct manipulation, describing how they would ``drag the nodes around to organize them the way I wanted.''
P3 spent significant time ``rearranging nodes and adding my own ideas because I wanted the graph to reflect how I think about the topic.''
Binks et al.~\cite{binks2022representational} found that users adopt diverse organizational strategies when structuring knowledge independently, and no single representational structure is universally effective for all purposes or ways of thinking.
The implication extends beyond the placement of individual nodes to the entire topology: the system's choice of hierarchy, grouping logic, and relationship types all encode organizational judgments that may diverge from a given user's mental model.
Future co-creation systems should treat AI-generated structure as provisional at every level---placement, grouping, hierarchy, and relationship types---and support user correction through both direct manipulation and natural language.
As interaction accumulates, systems could model individual users' organizational preferences to reduce the frequency of corrections over time, a form of computational Theory of Mind~\cite{li2023tom, street2024llmtheorymindalignment} applied to knowledge structure rather than dialogue.}
 
\remark{change log (Design Implications): (1)~DI1: Generalized cross-system contrast to all representational synchronization systems; added cross-representation preview recommendation. (2)~DI2: Condensed to focus on expansion granularity tension; added Sensecape comparison. (3)~DI3: Replaced original DI3 and closing Theory of Mind paragraph with new implication on provisional organizational structure, addressing R3 (representation choice) and R4 (no single correct representation). Added Binks et al.; relocated Theory of Mind as forward mechanism for reducing corrections.}

\subsection{Limitations and Future Work}
 
Our evaluation has limitations that suggest directions for future work.
The 12-participant study, while appropriate for formative evaluation of interactive systems~\cite{nielsen1994usability}, involved graduate students with prior AI experience and may not generalize to broader populations.
We measured user perceptions but did not conduct independent assessment of task outcomes, such as expert evaluation of slide deck quality.
Our pipeline evaluation focused on component-level accuracy rather than long-term system robustness as knowledge graphs grow larger; future work should conduct ablate the multi-agent architecture against simpler alternatives.
 
\revision{Our user study focused on slide deck preparation as a downstream task for knowledge construction.
While the task requires synthesizing, organizing, and presenting information from multiple sources, it represents one point in a broader space of knowledge work activities.
Preparing for an exam, conducting a literature review, or evaluating a new idea may require different balances of retrieval and contribution.
Future work should examine whether the contribution pathway's benefits generalize across these varied task contexts.}
 
\revision{Finally, \name{} represents knowledge as a hierarchical node-link graph, which is effective for modeling semantic and hierarchical relationships but may not support all organizational strategies equally well.
Future work should explore whether alternative visual representations, such as spatial clusters, matrix views, or timelines, might better match certain organizational preferences and complement the node-link graph with representational flexibility.}
 
\remark{change log (Limitations): Removed multimodal input, extended use, and collaborative scenarios sentences to tighten scope. Merged ablation studies point to pipeline robustness sentence. Added task scope paragraph (R3) and representation choice paragraph (R3, R4) as future directions.}
\section{Conclusion}

In this paper, we present \name{}, an interactive visual system to support human-AI collaborative knowledge construction, grounded by established principles in exploratory information seeking and knowledge externalization.
\revision{\name{} enables users to both query document-grounded information and contribute to the knowledge structure by adding concepts, modifying relationships, and reorganizing the hierarchy, producing a co-created knowledge graph where document-derived and user-contributed knowledge coexist.}
A multi-agent pipeline coordinates intent disambiguation, knowledge placement, and coherence maintenance, with each component validated through quantitative evaluation.
A user study with 12 participants compared \name{} against a retrieval-only baseline with graph visualization.
Overall, participants reported lower cognitive load and enhanced knowledge organization when exploring unfamiliar topics, particularly valuing the ability to progressively expand the knowledge graph and integrate their own insights into the evolving structure.

\begin{acks}
This work is supported in part by the Natural Sciences and Engineering Research Council of Canada (NSERC) Discovery Grant \#RGPIN-2020-03966, the Canada Foundation for Innovation (CFI) John R. Evans Leaders Fund (JELF) \#42371, and a gift fund from Adobe.
\end{acks}

\bibliographystyle{ACM-Reference-Format}
\bibliography{references.bib}

\clearpage

\appendix

\section{Prompts} \label{apx:prompts}
\subsection{Parse Query}

\begin{lstlisting}[basicstyle=\small\ttfamily, breaklines=true, frame=single]
You are a helpful assistant that classifies the user's query and generates an optimized query content.
When there's ambiguity in the user's query, use the chat history to infer the user's intention.
Ensure the query content is clear and can be executed without seeing the chat history.

There are three types of user's query:
1. Information Retrieval Queries:
   - Examples: "What is the definition of X?", "Is X the best player in the world?"
   - Generate query category: "search"
   - Generate query content: Preserve the original query without modifying meaning
2. Graph Editing Commands:
   - Examples: "Add a new concept called 'X' under the concept 'Y'"
   - Generate query category: "edit"
   - Generate query content: Preserve the original command without modification
3. Expansion Requests:
   - Examples: "Tell me more about X", "Explain X in more detail", "Elaborate on X"
   - Generate query category: "expansion"
   - Generate query content: "What are the sub-topics covered in the document related to 'X'?"

User: {query}
Context: {chat_history}
\end{lstlisting}

The Oracle Module uses this classification system to route queries to the appropriate processing pipeline and to reformat ambiguous queries for optimal processing.
   
\subsection{Update Graph}

\begin{lstlisting}[basicstyle=\small\ttfamily, breaklines=true, frame=single]
User's question: {query}
Answer: {response}

Update the graph based on the question and answer:

1. Create a new node for the answer if the answer does not fit under any existing node. 
   Link the new node to the existing nodes that are related to the answer.
   Consider both the node name and description when identifying related existing nodes.
   For example, if the user asks "What is the definition of XXX?", check if there's a node 
   with the name "XXX" or a description containing "XXX".

2. Break down the user's question and the answer into key points.

3. Maintain hierarchy: general key points as parent nodes, specific details as child nodes.

4. For each key point:
   a. Identify the relevant nodes in the graph related to the key point.
   b. If the key point is a sub-topic of an existing node, add it as a child node.
   c. If the key point is a parent-topic of an existing node, add it as a parent node.

Remember: You're creating a hierarchical knowledge graph, not a flat list.
\end{lstlisting}

This prompt guides the system in maintaining hierarchical relationships when updating the knowledge graph based on new information from user interactions.

\subsection{Query Graph}

\begin{lstlisting}[basicstyle=\small\ttfamily, breaklines=true, frame=single]
You are an assistant for question-answering tasks. You will be given a question and a context.
Use ONLY the following pieces of retrieved context to answer the question.
If you lack sufficient information from the context, respond with 'I don't know'.
Do not fabricate or assume any information not present in the context.

Your answer should resemble a hierarchical map, describing the relationships between each topic and the central keyword of the user's input.
For each topic, explain how it is related to the central keyword, using specific information from the context.

The context is: {context}
\end{lstlisting}

This prompt instructs the system to generate structured responses resembling a knowledge graph based strictly on the provided context, maintaining clear relationships between topics and the central concept.

\subsection{Refine Query}

\begin{lstlisting}[basicstyle=\small\ttfamily, breaklines=true, frame=single]
You are an intelligent query refiner. Your task is to analyze the user's original query and the response from the graph retriever, then generate a refined query for the RAG retriever.

Guidelines:
1. Focus on the parts of the query that cannot be answered by the graph.
2. Make the refined query more specific and targeted.
3. Remove any parts of the query that can already be answered by the graph.
4. Ensure the refined query is clear and self-contained.
5. If the entire query can be answered by the graph, generate a minimal query to confirm or expand on the information.
Directly output the refined query, do not output any other text.

Original query: {original_query}
Graph retriever response: {graph_response}
\end{lstlisting}

This prompt helps the system refine user queries by identifying information gaps in the current knowledge graph, ensuring that subsequent retrievals are targeted and non-redundant.

\subsection{Query Knowledge Base}

\begin{lstlisting}[basicstyle=\small\ttfamily, breaklines=true, frame=single]
You are an assistant for question-answering tasks. You will be given a question and a context.
Use ONLY the following pieces of retrieved context to answer the question.
If you lack sufficient information from the context, respond with 'I don't know'.
Do not fabricate or assume any information not present in the context.
Your answer should resemble a mind map, describing the relationships between each topic and the central keyword of the user's input.
For each topic, explain how it is related to the central keyword, using specific information from the context.

The context is: {context}
\end{lstlisting}

This prompt guides the system to generate structured responses organized as a knowledge graph that maintain clear relationships.

\subsection{Generate Suggestions}

\begin{lstlisting}[basicstyle=\small\ttfamily, breaklines=true, frame=single]
You are an intelligent agent responsible for generating suggestions for expanding a knowledge graph. Your task is to determine the most logical relationships between potential new content and one specific existing node.

1. Read the given existing node content carefully and understand the context.
2. Based on your knowledge and the context of the existing node, generate relevant suggestions for expansion.
3. Review the existing suggestions of the current node and do not suggest the same topic twice.
4. Determine the most appropriate relationship between the new content and the existing node.
5. Provide a list of suggestions, where each suggestion includes:
   - A topic that could be added as a child of the current node
   - A brief description of that topic as a full, informative sentence
   - The relationship between the new content and the existing node

Ensure that:
1. The suggestions are directly related to the existing node with a logical relationship.
2. Only include content that directly fits as children of the current node.
3. If you don't have any suggestions, just return an empty list.
4. Aim to provide 3-5 relevant and diverse suggestions for expanding the graph, DO NOT EXCEED 5.
5. The description should be a complete, informative sentence that addresses specific aspects or examples related to the topic.

Example of a good description:
Topic: 'Applications of Machine Ethics'
Description: 'Machine ethics is applied in various real-world scenarios, including autonomous vehicles making moral decisions in potential accident situations, AI systems in healthcare prioritizing patient care, and military AI navigating complex ethical dilemmas in combat situations.'

Existing nodes on the graph: {existing_nodes}
Expanding from node: {node_info}
Existing suggestions: {existing_suggestions}
Related content in the docs: {content}
\end{lstlisting}

This prompt guides the system in generating contextually relevant suggestions for expanding specific nodes in the knowledge graph, ensuring diversity and logical relationships.

\subsection{Rewrite Final Output}

\begin{lstlisting}[basicstyle=\small\ttfamily, breaklines=true, frame=single]
User's input: {input}
Refined query based on the input: {query}

Your task is to rewrite the following answer: {answer} such that the response answers both the user's input and the refined query.
After the final answer, add a reference section that lists the sources of the answer: {sources}
After the reference section, add a note to the user that the question is asked specifically on the node {node_name}

Rewriting guidelines:
1. Improve readability without modifying the content of the original answer.
2. The final answer should answer the user's input and the refined query.
3. Keep the answer concise and to the point.
4. Improve formatting for clarity if possible.
5. The final answer should introduce the main themes directly. Do not have text "High-level summary" or "Detailed bullet points" in the final answer.
\end{lstlisting}

This prompt instructs the system to produce concise, well-formatted responses that address user queries.


\begin{figure*}[t!]
    \section{Baseline User Interface} \label{apx:baseline}
    \centering
    \includegraphics[width=\linewidth]{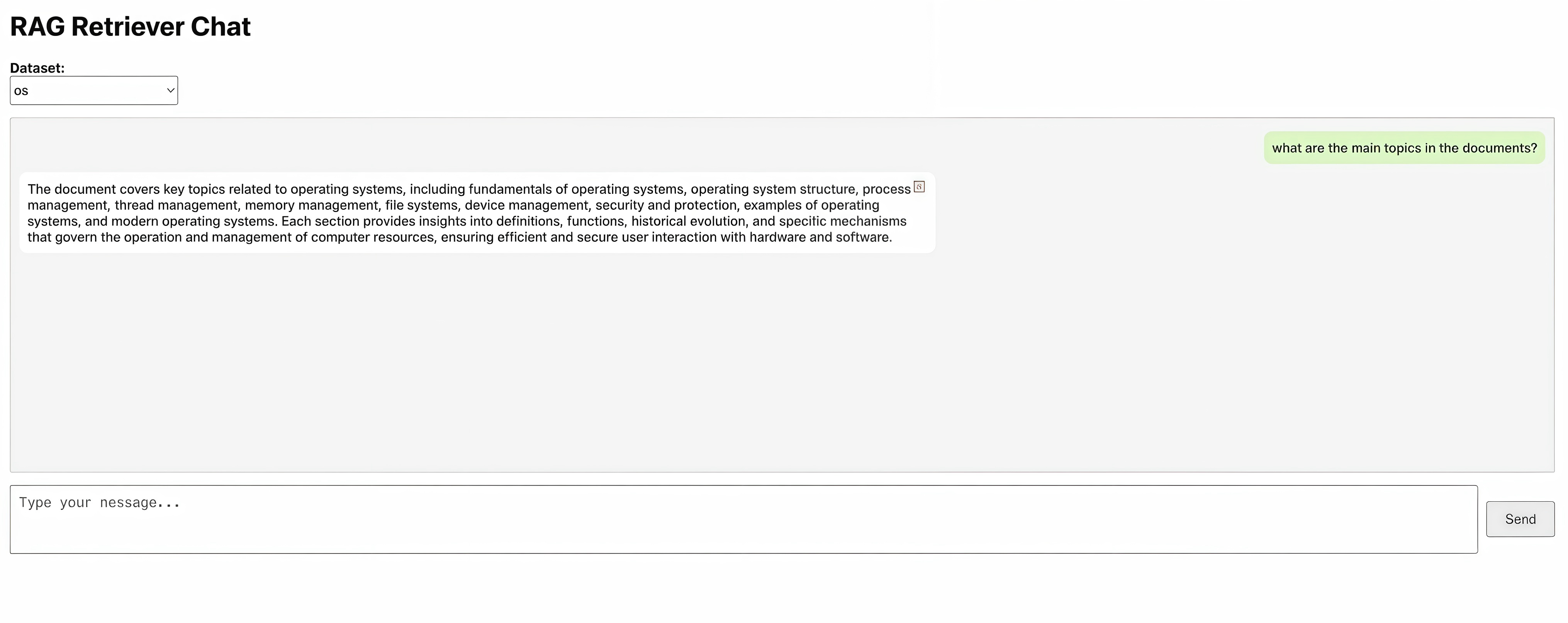}
    \vspace{-13mm}
    \caption*{(a) Baseline RAG Retriever Interface. Participants select a dataset from the dropdown menu and pose questions to receive relevant responses.}
    \vspace{4mm}
    \includegraphics[width=\linewidth]{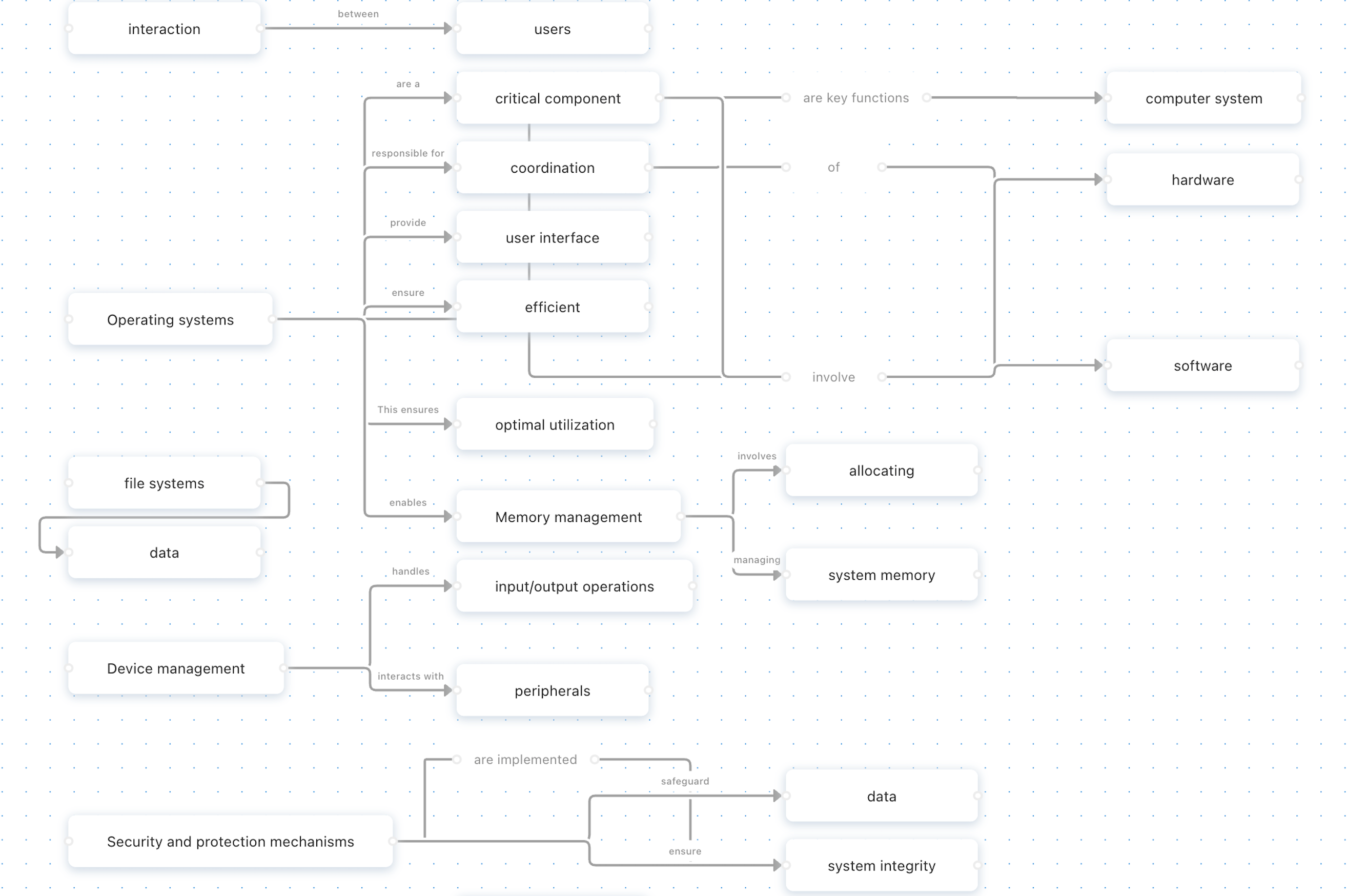}
    \vspace{-5mm}
    \caption*{(b) Baseline Visualization Interface. Participants input responses from the retriever and explore the knowledge base through automatically generated node-link diagrams.}
    \vspace{2mm}
    \caption{Baseline system interfaces used in the user study.}
    \Description{Two baseline interface screenshots stacked vertically. (a) A chat interface with a dataset dropdown and text input for posing questions. (b) An automatically generated node-link diagram about operating systems showing concepts like file systems, memory management, and device management connected by labeled edges.}
    \label{fig:baseline}
\end{figure*}
\end{document}